\documentclass[useAMS,usenatbib]{mn2e}
\input psfig.sty

\def \apj {ApJ}
\def \apjl {ApJL}
\def \mnras {MNRAS}
\def \apjs {ApJS}
\def \aap {A\&A}
\def \nat {Nature}
\def \etal {et~al.~}

\def \spose#1{\hbox  to 0pt{#1\hss}}
\def \lta{\mathrel{\spose{\lower 3pt\hbox{$\sim$}}\raise  2.0pt\hbox{$<$}}}
\def \gta{\mathrel{\spose{\lower  3pt\hbox{$\sim$}}\raise 2.0pt\hbox{$>$}}}

\def \kms {\ifmmode  \,\rm km\,s^{-1} \else $\,\rm km\,s^{-1}  $ \fi }
\def \kpc {\ifmmode  {\rm kpc}  \else ${\rm  kpc}$ \fi  }  
\def \Msun {\ifmmode M_{\odot} \else $M_{\odot}$ \fi} 
\def \Mstar {\ifmmode M_{\rm star} \else $M_{\rm star}$ \fi}
\def \Mgas {\ifmmode M_{\rm gas} \else $M_{\rm gas}$ \fi}

\def \LCDM {\ifmmode \Lambda{\rm CDM} \else $\Lambda{\rm CDM}$ \fi}
\def \Omegam {\ifmmode \Omega_{\rm m} \else $\Omega_{\rm m}$ \fi}
\def \Omegab {\ifmmode \Omega_{\rm b} \else $\Omega_{\rm b}$ \fi}
\def \OmegaL {\ifmmode \Omega_{\rm \Lambda} \else $\Omega_{\rm \Lambda}$\fi}

\def \Rvir {\ifmmode R_{\rm vir} \else $R_{\rm vir}$ \fi}
\def \Vvir {\ifmmode V_{\rm  vir} \else  $V_{\rm vir}$  \fi}
\def \Vmax {\ifmmode V_{\rm  max} \else  $V_{\rm max}$  \fi}
\def \Vlast {\ifmmode V_{\rm  last} \else  $V_{\rm last}$  \fi}
\def \Mvir {\ifmmode M_{\rm  vir} \else $M_{\rm  vir}$ \fi}
\def \Jvir {\ifmmode J_{\rm vir} \else $J_{\rm vir}$ \fi} 
\def \lam {\ifmmode \lambda  \else $\lambda$ \fi}
\def \lamp {\ifmmode \lambda^{\prime} \else $\lambda^{\prime}$  \fi}

\def \Mgal {\ifmmode M_{\rm gal} \else $M_{\rm gal}$ \fi}
\def \Jgal {\ifmmode J_{\rm gal} \else $J_{\rm gal}$ \fi}
\def \mgal {\ifmmode m_{\rm gal} \else $m_{\rm gal}$ \fi}
\def \rj {\ifmmode {\cal R}_j \else ${\cal R}_j$ \fi} 
\def \egf {\ifmmode {\epsilon}_{\rm GF} \else ${\epsilon}_{\rm GF}$ \fi} 
\def \efb {\ifmmode {\epsilon}_{\rm FB} \else ${\epsilon}_{\rm FB}$ \fi} 
\def \flost {\ifmmode {f}_{\rm lost} \else ${f}_{\rm lost}$ \fi} 
\def \lamgal {\ifmmode \lambda_{\rm gal} \else $\lambda_{\rm gal}$ \fi}
\def \lampgal {\ifmmode \lambda'_{\rm gal} \else $\lambda'_{\rm gal}$ \fi}
\def \Vcirc {\ifmmode V_{\rm circ} \else $V_{\rm circ}$ \fi}
\def \Vrot {\ifmmode V_{\rm rot} \else $V_{\rm rot}$ \fi}

\title[Angular momentum of disc galaxies]
      {The angular momentum of disc galaxies: implications for gas accretion, 
       outflows, and dynamical friction}

\author[A.A. Dutton \& F.C. van den Bosch] 
       {Aaron A. Dutton$^{1}$\thanks{dutton@uvic.ca}\thanks{CITA National Fellow},
        Frank C. van den Bosch$^{2}$\\
  $^1$Department of Physics and Astronomy, University of Victoria,
      Victoria, BC, V8P 5C2, Canada\\
  $^2$Astronomy Department, Yale University, P.O. Box 208101, New Haven, 
      CT 06520-8101, USA}

\begin{document}
             
\date{Accepted 2011 December 5. Received 2011 October 29; in original form 2011 June 2}
             
\pagerange{\pageref{firstpage}--\pageref{lastpage}}\pubyear{2012}

\maketitle           

\label{firstpage}
             

\begin{abstract}
  We combine constraints on the galaxy-dark matter connection with
  structural and dynamical scaling relations to investigate the
  angular momentum content of disc galaxies. For haloes with masses in
  the interval $10^{11.3}\Msun \lta \Mvir \lta 10^{12.7}\Msun$ we find
  that the galaxy spin parameters are basically independent of halo
  mass with $\langle \lambda'_{\rm gal} \rangle \equiv (\Jgal / \Mgal)
  / (\sqrt{2}\Rvir\Vvir) = 0.019^{+0.004}_{-0.003} \,(1\sigma)$. This
  is significantly lower than for relaxed \LCDM haloes, which have an
  average spin parameter $\langle \lambda'_{\rm halo} \rangle =
  0.031\pm 0.001$. The average ratio between the specific angular
  momentum of disc galaxies and their host dark matter haloes is
  therefore $\rj \equiv \lambda_{\rm gal}' / \lambda'_{\rm halo} =
  0.61^{+0.13}_{-0.11}$. This calls into question a standard
  assumption made in the majority of all (semi-analytical) models for
  (disc) galaxy formation, namely that $\rj=1$. Using simple disc
  formation models we show that it is particularly challenging to
  understand why $\rj$ is independent of halo mass, while the galaxy
  formation efficiency ($\egf$, proportional to the ratio of galaxy
  mass to halo mass) reveals a strong halo mass dependence. We argue
  that the empirical scaling relations between $\egf$, $\rj$ and halo
  mass require both feedback (i.e., galactic outflows) and angular
  momentum transfer from the baryons to the dark matter (i.e.,
  dynamical friction). Most importantly, the efficiency of angular
  momentum loss need to decrease with increasing halo mass. Such a
  mass dependence may reflect a bias against forming stable discs in
  high mass, low spin haloes or a transition from cold-mode accretion
  in low mass haloes to hot-mode accretion at the massive
  end. However, current hydrodynamical simulations of galaxy
  formation, which should include these processes, seem unable to
  reproduce the empirical relation between $\egf$ and $\rj$.  We
  conclude that the angular momentum build-up of galactic discs
  remains poorly understood.
\end{abstract}


\setcounter{footnote}{1}


\section{Introduction}
\label{sec:intro}

In the current paradigm of galaxy formation, galaxy discs are
considered to form from the accretion of gas inside hierarchically
growing cold dark matter (CDM) haloes (see Mo, van den Bosch \& White
2010 for a comprehensive overview). The dark matter (DM) and gas
acquire angular momentum via tidal torques in the early Universe
(Peebles 1969). When gas accretes onto the central galaxy, this
angular momentum may eventually halt the collapse and lead to the
formation of a rotationally supported disc (Fall \& Efstathiou 1980).
If the specific angular momentum of the baryons is conserved during
galaxy formation, there is enough angular momentum to make galaxy
discs with the observed sizes (e.g., Dalcanton \etal 1997; Mo, Mao, \&
White 1998; Firmani \& Avila-Reese 2000; de Jong \& Lacey 2000;
Navarro \& Steinmetz 2000). In fact, models in which galaxies have
$\sim 30\%$ lower specific angular momentum than their dark matter
haloes may be preferred (Dutton \etal 2007).  Another success of this
picture is that the large variation in the specific angular momentum
of dark matter haloes, at fixed halo mass, explains the large scatter
in disc sizes, or equivalently surface brightnesses, at fixed
luminosity (Dalcanton \etal 1997). This picture also accounts for the
time evolution of disc sizes (Somerville \etal 2008; Firmani \&
Avila-Reese 2009; Dutton \etal 2011a), reproducing the observed weak
evolution since redshift $z\simeq 1$ (Barden \etal 2005), and the
stronger evolution observed from $z\sim 2$ to $z\sim 1$ (Trujillo
\etal 2006; Williams \etal 2010).

Despite the many successes of these models, all of which assume that
specific angular momentum is conserved during galaxy formation, there
are several reasons, from a physical point of view, as to why the
ratio between the specific angular momentum of the baryons and that of
the dark matter halo, denoted $\rj$\footnote{Note that in the notation
  of Mo, Mao, \& White (1998), $\rj=(j_{\rm d}/m_{\rm d})$, where
  $j_{\rm d}$ is the ratio between the {\it angular momentum} of the
  disc and dark matter halo, and $m_{\rm d}$ is the ratio between the
  {\it mass} of the disc and dark matter halo.}, might be different
from unity:

\begin{enumerate}

\item Selective accretion of baryons. If not all available baryons
  make it into the galaxy, the specific angular momentum of the galaxy
  can be either higher ($\rj > 1$) or lower ($\rj < 1$) than that of
  all baryons (whether it is higher or lower simply depends on the
  specific angular momentum distribution of the materials that {\it
    are} selected). An example of a selective accretion process is
  cooling. Cooling is generally an inside-out process, with the inner
  regions cooling prior to the more outer regions. Since the mass of
  \LCDM haloes is more concentrated than the specific angular momentum
  (see e.g., Navarro \& Steinmetz 1997), one expects that $\rj < 1$ if
  not all baryons manage to cool.

\item Exchange of angular momentum between the dark matter and the
  baryons.  An example of this is dynamical friction, which transfers
  orbital angular momentum from the baryons to the dark matter, thus
  resulting in $\rj < 1$ (e.g., Navarro \& Benz 1991; Navarro \& White
  1994). The strength of this process depends on the efficiency of
  star formation at early times. If the formation of baryonic clumps
  can be suppressed, such as through strong feedback, then the angular
  momentum transfer is reduced (e.g., Weil, Eke \& Efstathiou 1998;
  Sommer-Larsen, Gelato \& Vedel 1999; Eke, Efstathiou \& Wright
  2000).

\item Feedback. If feedback (from supernova, AGN or any other
  mechanism) results in the removal of gas, one can achieve values for
  $\rj$ that are very different from unity by simply removing a
  specific subset of the baryons. As an example, several studies have
  argued that supernova feedback affects low angular momentum material
  more than high angular momentum material, causing a net increase of
  $\rj$ (Maller \& Dekel 2002; Dutton \& van den Bosch 2009; Dutton
  2009; Governato \etal 2010; Brook \etal 2011).

\end{enumerate}

Cosmological hydrodynamical simulations almost invariably predict that
disc galaxies have significantly less specific angular momentum than
their dark matter haloes ($\rj < 1$), suggesting that mechanisms (i)
and (ii) discussed above play an important role.  Early simulations in
a standard cold dark matter (SCDM) cosmology (i.e. $\Omegam=1,
\OmegaL=0$) resulted in a catastrophic loss of angular momentum
(Navarro \& Benz 1991; Navarro \& White 1994; Steinmetz \& Navarro
1999), producing discs an order of magnitude too small. More recent
simulations in $\Lambda$ Cold Dark Matter ($\LCDM$) cosmologies have
resulted in less angular momentum loss, but typical ratios between the
specific angular momentum of the baryons and that of the dark matter
are $\rj \simeq 50\%$ (Piontek \& Steinmetz 2011; Sales \etal 2010).
This ``angular momentum loss'' is sometimes attributed to lack of
numerical resolution (e.g. Governato \etal 2004; Kaufmann \etal 2007).
However, Piontek \& Steinmetz (2009) argue that the dominant cause is
dynamical friction between clumps of baryons and the dark matter halo.
If, based on these findings, we assume that angular momentum loss is a
natural feature of disc galaxy formation in a \LCDM universe, a
relevant question is the following: {\it If the baryons that make up a
  discy galaxy carry half the specific angular momentum of the dark
  matter halo in which they reside, is it possible to reproduce in
  detail the observed sizes of disc galaxies}?

This question was addressed by Navarro \& Steinmetz (2000).  Using
scaling relations between disc size, luminosity, and rotation
velocity, these authors derived an expression for $\rj$ ($f_j$ in
their notation) that depends on the galaxy formation efficiency, $\egf
\equiv (\Mgal/\Mvir)/(\Omegab/\Omegam)$ ($f_{\rm bdsk}$ in their
notation), cosmic matter density, $\Omegam$, and stellar mass-to-light
ratio, $\Upsilon_{I}$.  They found that if the rotation speeds of
galaxy discs are approximately the same as the circular velocities of
their surrounding haloes (which has subsequently been shown to be the
case, at least on average, by Dutton \etal 2010b), then discs must
have retained about half of the available specific angular momentum
during their assembly. Another result from this analysis is that disc
galaxies forming in low density cosmogonies, such as $\LCDM$, need to
have $\rj \simeq 2 \,\egf$, that is, they draw a larger fraction of
the specific angular momentum than the galaxy formation efficiency.
This is at odds with the expectation (see \S~\ref{sec:cooling} below)
that $\rj \simeq \egf$ if the galaxy formation efficiency is mainly
determined by the inefficiency of cooling.

Thus observational measurements of the specific angular momentum in
galaxies place constraints on the way galaxies acquire their mass and
angular momentum.  The specific angular momentum of the dark matter,
$\Jvir/\Mvir$, where $\Jvir$ is the total angular momentum, and
$\Mvir$ is the total virial mass, is commonly expressed in terms of
the dimensionless spin parameter (Peebles 1969):
\begin{equation}
\label{eq:spin}
\lambda_{\rm halo} = {J_{\rm vir} \vert E \vert^{1/2} \over G M_{\rm
    vir}^{5/3}} = \frac{\Jvir/\Mvir}{\sqrt{2}\, \Rvir \Vvir} \,
f_{c}^{1/2} = \lambda^{\prime}_{\rm halo} \, f_{c}^{1/2}
\end{equation}
Here $E$ is the halo's energy, $\Rvir$ is the virial radius, $\Vvir$
is the circular velocity at the virial radius, and $f_c$ measures the
deviation of $E$ from that of a singular isothermal sphere truncated
at $\Rvir$.  It is common to set $f_c=1$ (Bullock \etal 2001), so that
the spin parameter only depends on the total angular momentum and mass
of the halo.  In this case the spin parameter is denoted $\lamp_{\rm
  halo}$. The equivalent spin parameter for the galaxy is defined as
\begin{equation}\label{eq:lamprime}
  \lambda^{\prime}_{\rm gal} = {(J_{\rm gal}/M_{\rm gal})  \over 
\sqrt{2} \Rvir \Vvir}\,,
\end{equation}
(see Section~\ref{sec:def} for details).

Dark matter only cosmological simulations have shown that both
$\lambda_{\rm halo}$ and $\lamp_{\rm halo}$ are log-normally
distributed (e.g., Bullock \etal 2001) with median and scatter
independent of halo mass (e.g., Macci\`o \etal 2007; Bett \etal 2007;
Mu\~noz-Cuartas \etal 2011) and independent of environment (Lemson \&
Kauffmann 1999, Macci\`o \etal 2007).  For relaxed haloes identified
with a spherical overdensity (SO) algorithm Macci\`o \etal (2008) find
a median value of $\lamp_{\rm halo} = 0.031\pm0.001$.  Using a similar
halo definition Bett \etal (2007) find a median $\lambda_{\rm
  halo}=0.0367$.  For the cosmology adopted by Bett \etal, $f_c\simeq
1.3$ for a $10^{12}\Msun$ halo, and thus $\lambda^{'}_{\rm halo}
\simeq 0.032$, which is in excellent agreement with the results of
Macci\`o \etal (2008).
Simulations with gas in which there is no cooling find similar spin
parameters for the baryons and dark matter (van den Bosch \etal 2002;
Sharma \& Steinmetz 2005). Thus in what follows we assume that the
median spin parameter of the halo is
\begin{equation}
  \lambda^{'}_{\rm halo}=0.031\pm0.001.
\end{equation}

The main goal of this paper is to determine the average value of $\rj$
for a large sample of disc galaxies as a function of their halo mass,
and to use semi-analytical models to explore implications for how disc
galaxies form. Since it is impossible to measure the specific angular
momentum (or spin parameter) of a real (as opposed to simulated) dark
matter halo, one can only obtain an estimate for the {\it average}
value of $\rj$ by comparing the {\it distribution} of galaxy spin
parameters, $\lamp_{\rm gal}$, obtained from a sample of galaxies to
the {\it distribution} of halo spin parameters, $\lamp_{\rm halo}$,
obtained from numerical simulations of structure formation in a \LCDM
cosmology. This is the approach we will follow in this paper.

Measuring $\lamp_{\rm gal}$ for an individual galaxy necessarily
involves the following steps:
\begin{enumerate}
\item Determine the specific angular momentum of the baryons. 
\item Determine the mass (and virial radius) of the dark matter halo
  in which the galaxy resides. 
\end{enumerate}
Step (i) is relatively straightforward. Using that, for a disc galaxy,
\begin{equation}
\label{eq:jgal}
J_{\rm gal} = 2 \pi \int_0^{\Rvir} \Sigma(R) \, V_{\rm rot}(R) \, R \, {\rm d}R\,,  
\end{equation}
it is clear that one can determine the angular momentum of a disc
galaxy from measurements of its surface mass density, $\Sigma(R)$,
(accounting for both gas and stars), and its rotation curve $V_{\rm
  rot}(R)$ (see e.g., van den Bosch, Burkert \& Swaters
2001). Alternative, since disc galaxies in general have exponential
surface density profiles and flat rotation curves, one can use simple
scaling relations between mass (or luminosity), size and
characteristic rotation velocity in order to obtain an estimate of
$J_{\rm gal}$ without having to measure $\Sigma(R)$ or $V_{\rm
  rot}(R)$ (e.g., Navarro \& Steinmetz 2000; Tonini \etal 2006;
Hernandez \& Cervantes-Sodi 2006; Hernandez \etal 2007; Berta \etal
2008). Although less accurate than using Eq.~(\ref{eq:jgal}), this
method has the advantage that it can be applied to much larger samples
of galaxies.  Step (ii) is a little bit more involved. Typically,
accurate measurements of halo mass require probes on scales of the
virial radius. Examples of such `probes' are gravitational lensing,
X-ray data and satellite kinematics. However, unless the galaxy in
question is the central galaxy of a cluster-sized halo, such
measurements are rarely available for individual systems, and one has
to rely on uncertain estimates of the halo mass from the galaxy's
rotation curve. Alternatively, one can use various statistical
techniques to assign to each galaxy an {\it average} halo mass. Recent
years have seen dramatic progress in using a variety of techniques to
constrain the average relation between galaxy properties (typically
luminosity of stellar mass) and halo mass, in particular from galaxy
clustering (e.g., Yang, Mo \& van den Bosch 2003; Tinker \etal 2005),
galaxy-galaxy lensing (e.g, Mandelbaum \etal 2006; Cacciato \etal
2009), abundance matching (e.g., Conroy, Wechsler \& Kravtsov 2006),
and satellite kinematics (e.g., More \etal 2009).

In this paper we compute the average specific angular momentum and
spin parameters of disc galaxies as a function of their halo mass.  We
improve on previous studies in a number of ways. We constrain halo
masses using recent results for the halo mass - stellar mass relation
for disc galaxies obtained from weak gravitational lensing and
satellite kinematics (Dutton \etal 2010b), rather than abundance
matching results as in Tonini \etal (2006), or the highly questionable
assumption of a constant galaxy mass fraction made by Hernandez \&
Cervantes-Sodi (2006), Hernandez \etal (2007), and Berta \etal (2008).
We also improve upon these studies by taking account of realistic
uncertainties on halo masses.  We consider both stellar and gaseous
discs, rather than just stellar discs, which is important for low mass
galaxies which tend to be gas rich (Tonini \etal 2006). We calculate
disc angular momentum by integrating a model that is constructed to
reproduce the size-mass and velocity-mass (i.e., Tully-Fisher, Tully
\& Fisher 1977) relations, rather than simple scaling arguments for
exponential discs in isothermal haloes (i.e., $J_{\rm gal}/M_{\rm
  gal}= 2 R_{\rm d} V_{\rm rot}$), as used by e.g., Navarro \&
Steinmetz (2000) and Hernandez \etal (2007). We also discuss how
uncertainties related to stellar mass-to-light ratios and adiabatic
contraction impact the inferred values of $\rj$, and use simple models
of disc galaxy formation to discuss the implications of our findings
for cooling, outflows and angular momentum loss.

We adopt a flat \LCDM cosmology with matter density parameter
$\Omegam=0.27$, baryon density $\Omegab=0.044$, and Hubble parameter
$h=H_0/100 {\rm km\, s^{-1}\, Mpc^{-1}} = 0.7$. Dark matter halo
masses are defined so that the mean density of the halo (assumed to be
spherical) is 200 times the critical density.


\section{Definitions and Mass Models}
\label{sec:methods}

This section gives a brief overview of the main parameters we discuss
in this paper, and the constrained mass models we use to constrain
them.

\subsection{Mass and Angular Momentum Parameters}
\label{sec:def}

There are four main parameters that are relevant for what follows:
galaxy mass, $\Mgal$, halo virial mass, $\Mvir$, galaxy angular
momentum, $\Jgal$, and the angular momentum of the dark matter halo,
$\Jvir$. The galaxy mass is the sum of the mass in stars and cold
gas. The virial mass is the total mass (stars, cold gas, hot gas, dark
matter) within the virial radius.  The galaxy angular momentum is the
sum of the angular momenta of the stars and of the cold gas. The
virial angular momentum is the total angular momentum within the
virial radius.

Our main goal is to constrain the angular momentum ratio, $\rj$,
defined as the ratio between the galaxy specific angular momentum and
the total specific angular momentum\footnote{Note that $\rj$ is not an
  efficiency because it can be larger than unity.}:
\begin{equation}\label{eq:rj}
  \rj \equiv {(J_{\rm gal}/M_{\rm gal}) \over (J_{\rm vir}/M_{\rm vir})} =
{\lamp_{\rm gal} \over \lamp_{\rm halo}}\,,
\end{equation}
where the spin parameters $\lamp_{\rm halo}$ and $\lamp_{\rm gal}$ are
defined by Eqs.~(\ref{eq:spin}) and~(\ref{eq:lamprime}), respectively.
Using that for disc galaxies $J_{\rm gal} \propto R_{\rm d} \, M_{\rm
  gal} \, V_{\rm rot}$, where $R_{\rm d}$ is the disc scale length
and $V_{\rm rot}$ is a characteristic rotation velocity of the disc,
one can use Eq.~(\ref{eq:rj}) to show that
\begin{equation}\label{eq:rd}
R_{\rm d} \propto \rj \, \lamp_{\rm halo} \, \left({V_{\rm rot} \over 
V_{\rm vir}}\right)^{-1} \, R_{\rm vir} \,, 
\end{equation}
which is the standard approach used in analytical and semi-analytical
models to predict the sizes of disc galaxies (e.g., Kauffmann 1996;
Mo, Mao \& White 1998; Somerville \& Primack 1999; van den Bosch 2000;
Cole \etal 2000). In many cases these studies simplify
Eq.~(\ref{eq:rj}) by assuming that $\rj = 1$ and that $V_{\rm
  rot}/V_{\rm vir}$ is a constant.  In this paper we examine whether
there is empirical support for the former assumption and what this
implies for the formation of disc galaxies in general. As we will
demonstrate, it is instructive to compare $\rj$ to the galaxy
formation efficiency parameter
\begin{equation}\label{eq:egf}
\egf \equiv {(\Mgal/\Mvir) \over (\Omegab/\Omegam)}\,.
\end{equation}
which describes what fraction of the cosmologically available baryons
have ended up in a galaxy.

\begin{figure*}
\centerline{
\psfig{figure=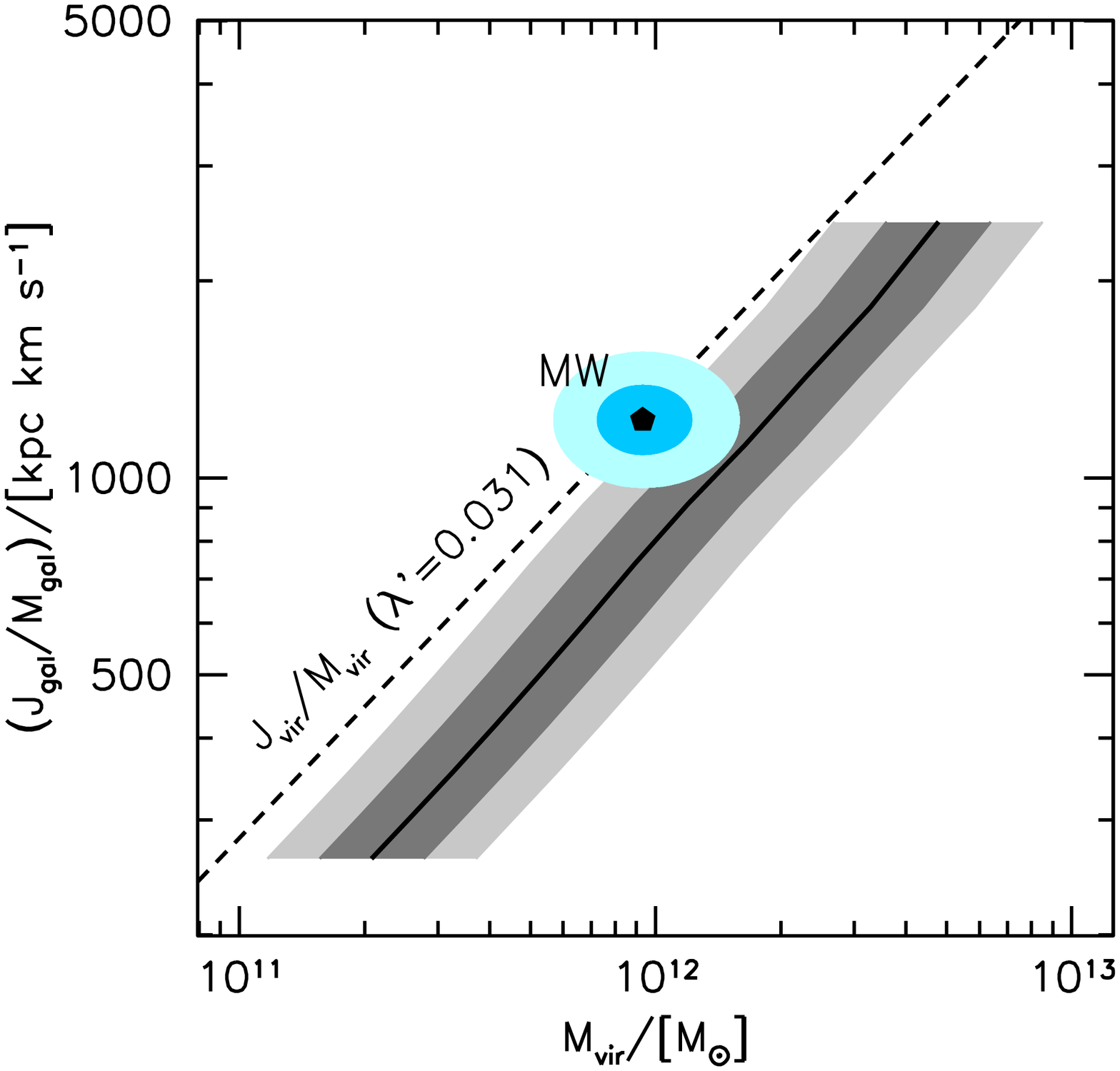,width=0.47\textwidth}
\psfig{figure=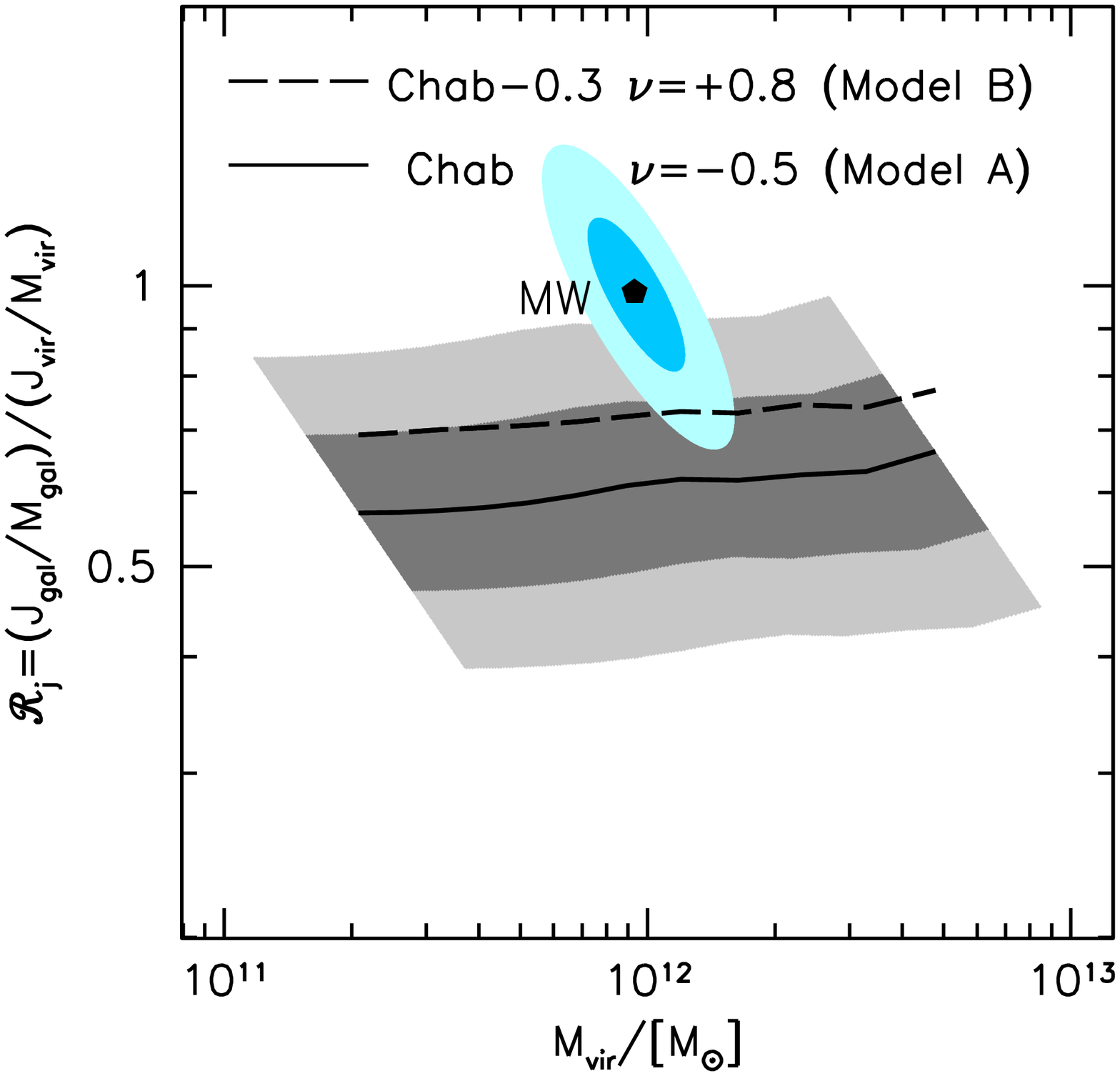,width=0.47\textwidth}
}
\caption{ {\it Left:} Galaxy specific angular momentum {\it vs.}
  virial mass for disc galaxies. The shaded regions indicate the one
  and two $\sigma$ uncertainties on virial mass.  The dashed line
  shows the relation between virial specific angular momentum and
  virial mass from cosmological N-body simulations, corresponding to a
  spin parameter $\lambda' = 0.031$.The pentagon shows estimates
    for the Milky Way (MW) with the shaded region showing one and two
    $\sigma$ uncertainties. {\it Right:} Specific angular momentum
  ratio ($\rj$) {\it vs.} virial mass. The solid line and shaded
  regions indicate the mean and the one and two $\sigma$ uncertainties
  for our fiducial model~A, which assumes a Chabrier (2003) IMF and a
  halo contraction model characterized by $\nu = -0.5$. The dashed
  line shows the mean for model~B, in which the stellar mass-to-light
  ratios are 0.3 dex smaller and $\nu = 0.8$. Note that, for both
  models, $\rj$ is less than unity and approximately independent of
  virial mass.}
\label{fig:jmvir}
\end{figure*}

\subsection{Mass Models}
\label{sec:mm}

We compute the {\it average} value for $\rj$ for disc galaxies using
the mass models of Dutton \etal (2011b), which have been constructed
to reproduce the observed structural and dynamical scaling relations
of late-type (i.e., star-forming and disc-dominated) galaxies. They
consist of four components: a stellar bulge, a stellar disc, a gas
disc, and a dark matter halo. Both the bulge and the dark matter halo
are assumed to be spherical and to follow a density distribution that
is given by a Hernquist profile (Hernquist 1990) and a NFW profile
(Navarro, Frenk, \& White 1997), respectively. The density profile of
the halo, however, is modified to account for the gravitational
response to galaxy formation (most often modelled as adiabatic
contraction, see Blumenthal et al. 1986). The disc is assumed to have
an exponential surface density and to be infinitesimally thin.

Each halo-galaxy system is described by nine parameters: three for the
halo (virial mass, $\Mvir$, concentration, $c$, and a parameter that
describes the impact of galaxy formation on the halo density profile);
and six for the baryons (stellar mass, $\Mstar$, gas mass, $M_{\rm
  gas}$, bulge fraction, $f_{\rm b}$, bulge size, $R_{\rm b}$, stellar
disc size, $R_{\rm d}$, and gas disc size, $R_{\rm g}$). The 6
parameters for the baryons are constrained by the observed scaling
relations: $R_{\rm b}$ {\it vs.} $\Mstar$; $R_{\rm d}$ {\it vs.}
$\Mstar$; $R_{\rm g}/R_{\rm d}$ {\it vs.} $\Mstar$; $f_{\rm b}$ {\it
  vs.} $\Mstar$, and $\Mgas/\Mstar$ {\it vs.} $\Mstar$ (see Dutton
\etal 2011b for details).  For the relation between $\Mvir$ and
$\Mstar$ we use the empirical relation from Dutton \etal (2010b),
which is inferred from a large number of independent studies and
techniques, including galaxy-galaxy lensing, satellite kinematics,
galaxy clustering, and abundance matching.  We will use the scatter
among these different studies/techniques, as an estimate for the
uncertainty in the $\Mvir$ {\it vs.} $\Mstar$ relation, and propagate
these uncertainties to those on $\egf$ and $\rj$. Finally, the
relation between halo mass, $\Mvir$, and halo concentration, $c$, is
taken from Macci\`o \etal (2008), and is calibrated using
high-resolution numerical simulations.

The usage of all these empirical and theoretical scaling relations
implies that the nine model parameters can be reduced to two primary
unknowns: the stellar initial mass function (IMF), and a prescription
for how galaxy growth influences the structural properties of the dark
matter halo (hereafter ``the halo contraction model''). It is common
practice to assume that disc formation is an adiabatic process and
that the halo responds by contracting using the adiabatic contraction
(hereafter AC) prescription of Blumenthal \etal (1986; hereafter
BFFP). However, numerical simulations have shown that this
prescription may not be sufficiently accurate, and several
alternatives have been proposed (e.g., Gnedin \etal 2004; Abadi \etal
2010). We follow Dutton \etal (2007) and use a generalized halo
contraction model, according to which a shell of dark matter initially
(before disc formation) at radius $r_i$ ends up, after disc formation,
at a radius
\begin{equation}\label{eq:ac}
r_f = \Gamma_{\rm BFFP}^{\nu} \, r_i\,.
\end{equation}
Here $\Gamma_{\rm BFFP}$ is the amount of contraction as predicted by
the AC-model of BFFP, and $\nu$ is a free parameter. Note that $\nu =
1$ corresponds to the standard adiabatic contraction model of BFFP,
while the Gnedin \etal (2004) and Abadi \etal (2010) models can be
well approximated by $\nu = 0.8$ and $\nu = 0.4$, respectively.  A
model without contraction has $\nu = 0$, while $\nu < 0$ indicates
that the dark matter halo responds to disc formation by expanding.

As shown in Dutton \etal (2011b), reproducing the observed
Tully-Fisher relation implies different values for $\nu$ for a
different choice of the IMF. In this paper we will consider two
different models that fit the Tully-Fisher relation (and all other
scaling relations mentioned above) equally well: model~A which adopts
a Chabrier (2003) IMF and $\nu = -0.5$ (i.e., halo expansion), and
model~B, in which we adopt the AC model of Gnedin \etal (2004), i.e.,
we set $\nu = 0.8$, and we assume an IMF that yields stellar
mass-to-light ratios that are 0.3 dex lower than for the Chabrier
(2003) IMF. As shown in Dutton \etal (2011b) both these models are in
excellent agreement with the data.  As we will show below, model~B
results in galaxy spin parameters that are $\sim 30$ percent ($\sim
0.12$ dex) higher than those of model~A.  This uncertainty is
comparable to that arising from the uncertainty in the $\Mvir$ {\it
  vs.} $\Mstar$ relation. Throughout we will mainly present the
results for model~A, while occasionally showing, for comparison, the
results for model~B.

For a given choice of IMF and $\nu$, the model completely specifies
the mass distributions of disc-plus-bulge galaxies as a function of
halo mass. From these mass models we compute the angular momentum of
the baryons by evaluating Eq.~\ref{eq:jgal}, where the surface density
is the sum of the surface densities of the gas and stars:
$\Sigma(R)=\Sigma_{\rm gas}(R)+\Sigma_{\rm star}(R)$.  We assume that
the bulge has no net angular momentum. If we assumed that bulges
follow the same rotation curve as the disc, then this would increase
the specific angular momentum of the galaxy by $\lta 20\%$. The effect
is largest for the most massive galaxies, but since massive bulges, in
general, are not rotationally supported, we believe our assumption
that bulges have no angular momentum to be of no major concern for
what follows.

A shown by Tonini \etal (2006) a significant fraction of the angular
momentum of a galaxy is in the cold gas. In our models the cold gas in
high mass galaxies ($M_{\rm gal}\simeq 10^{11}\Msun$) contains $\simeq
30\%$ of the angular momentum.  This fraction rises to over 50\% for
galaxy masses below $M_{\rm gal} \simeq 10^{10}\Msun$. Thus if cold
gas fractions increase at higher redshifts, as is commonly thought
(but see Dutton \etal 2010a), measuring angular momentum of high
redshift galaxies will require measurements of the distribution of
cold gas, and not just the stars.


\begin{figure*}
\centerline{
\psfig{figure=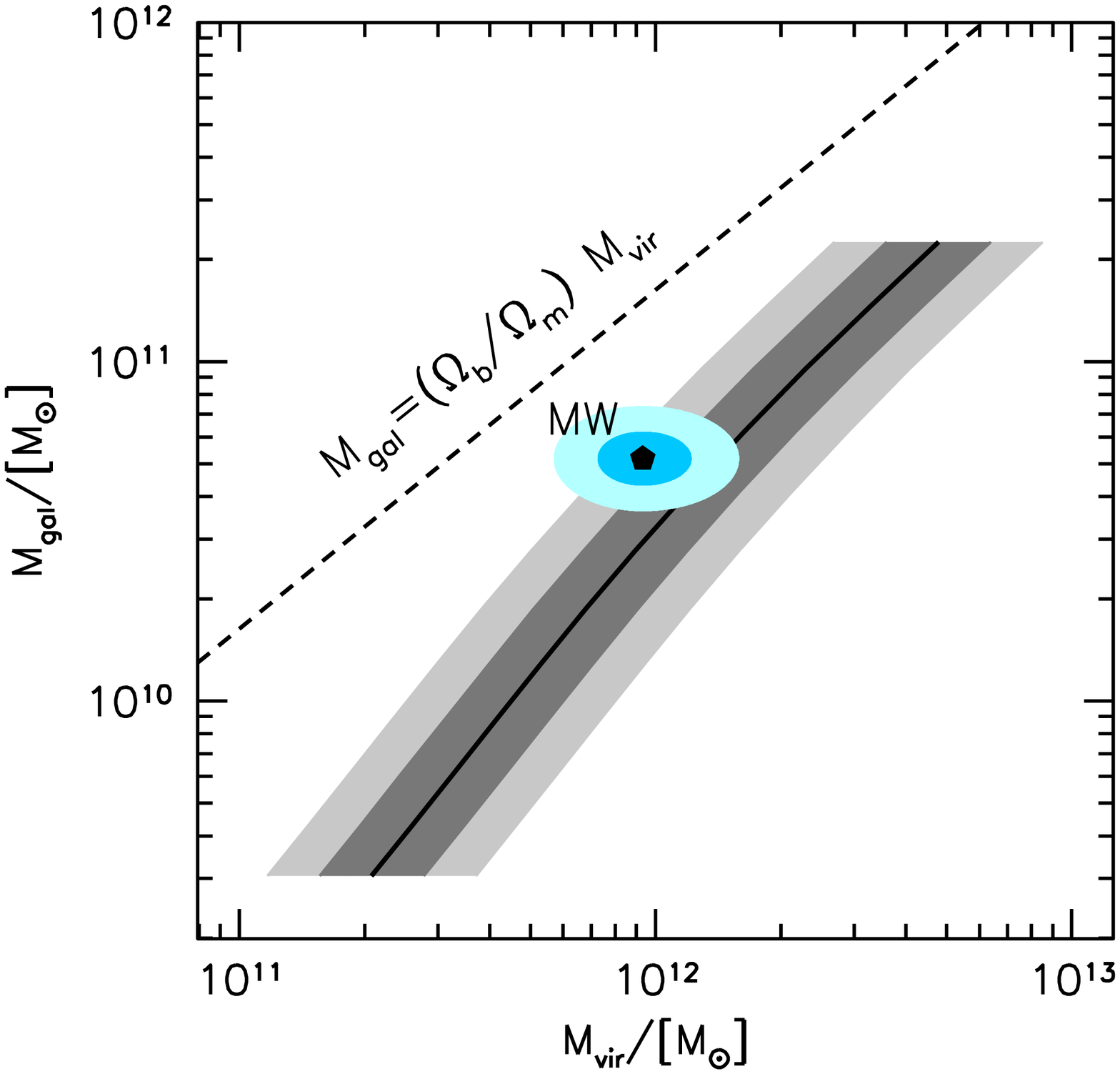,width=0.47\textwidth}
\psfig{figure=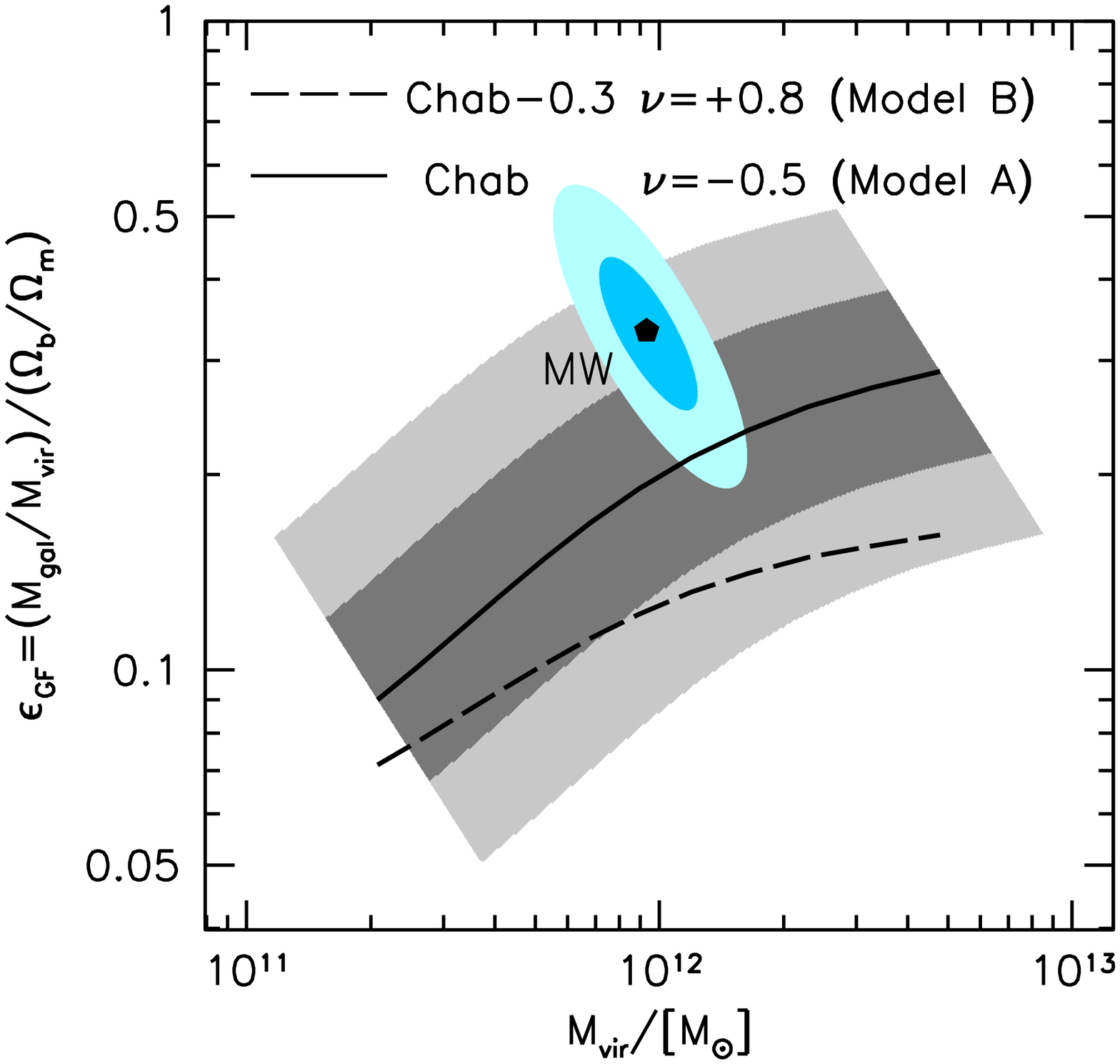,width=0.47\textwidth}
}
\caption{{\it Left:} Galaxy mass {\it vs.} virial mass for disc
  galaxies. The shaded regions reflect the one and two $\sigma$
  uncertainties on virial masses. The dashed line corresponds to the
  universal baryon fraction ($\Omegab/\Omegam=0.163$). The
    pentagon shows estimates for the Milky Way (MW), with the shaded
    region showing one and two $\sigma$ uncertainties. {\it Right:}
  Galaxy formation efficiency ($\egf$) {\it vs.} virial mass.  As in
  Fig.~\ref{fig:jmvir}, we show the results for both models A and
  B. For clarity, we only show the impact of virial mass uncertainties
  for our fiducial model A. Note that unlike $\rj$, which reveals no
  significant dependence on halo mass (see Fig.~\ref{fig:jmvir}), the
  galaxy formation efficiency varies by a factor of $\sim 3$ over the
  mass range probed.}
\label{fig:mmvir}
\end{figure*}

\section{Results}
\label{sec:results}

Using our constrained mass models we now discuss a number of
correlations between mass and angular momentum of disc galaxies and
dark matter haloes, and compare with previous results.  These
  correlations are presented in Figs.~\ref{fig:jmvir}-\ref{fig:spin}
  and Table~\ref{tab:data}.

\subsection{Galaxy Mass and Specific Angular Momentum}

The left-hand panel of Fig.~\ref{fig:jmvir} shows the correlation
between galaxy specific angular momentum and virial mass for
model~A. The shaded regions show the one and two $\sigma$
uncertainties in virial masses at fixed galaxy mass\footnote{Note that
  these are the uncertainties on the mean; they do not reflect the
  scatter.}, which is the dominant source of uncertainty.  The dashed
line shows the relation between virial specific angular momentum and
virial mass from cosmological N-body simulations (Macci\`o \etal
2008), which corresponds to a spin parameter of $\lambda'=0.031$.  The
galaxy specific angular momentum scales with virial mass as
$\Jgal/\Mgal \propto \Mvir^{0.70}$, which is very similar to the
virial scaling of $\Jvir/\Mvir \propto \Mvir^{0.67}$.
Thus the ratio $\rj$ between the galaxy and virial specific angular
momentum is roughly constant (right-hand panel of
Fig.~\ref{fig:jmvir}). However, the galaxy specific angular momentum
is significantly lower than the virial specific angular momentum with
$\rj = 0.61^{+0.13}_{-0.11}$ ($1\sigma$) for a virial mass of
$\Mvir=10^{12}\Msun$.  The dashed line in the right-hand panel of
Fig.~\ref{fig:jmvir} corresponds to model~B, and is shown for
comparison. Note that the combination of halo contraction and lower
stellar mass-to-light ratios result in $\rj$ values that are $\sim
0.1$ higher than for our fiducial model~A. This difference is
comparable to the 1$\sigma$ uncertainty on $\rj$ due to the
uncertainties in the relation between stellar mass and halo mass.

The left-hand panel of Fig.~\ref{fig:mmvir} shows the correlation
between galaxy mass and virial mass for the disc galaxies considered
here.  The dashed line shows the relation for the cosmically available
baryonic mass, i.e., $\Mgal = (\Omegab/\Omegam)\Mvir$, corresponding
to $\egf=1$.  The right-hand panel of Fig.~\ref{fig:mmvir} shows the
galaxy formation efficiency, $\egf$, which is the ratio between the
solid and dashed lines in the left-hand panel. Unlike the angular
momentum ratio, the galaxy formation efficiency is a strong function
of virial mass, varying by a factor of $\sim 3$ from $\egf\simeq 0.1$
at the low mass end ($\Mvir \sim 2 \times 10^{11}\Msun$) to $\egf
\simeq 0.3$ at the high mass end ($\Mvir \sim 5 \times
10^{12}\Msun$). Once again, the dashed line in the right-hand panel
corresponds to model~B, which yields lower galaxy formation
efficiencies due to the lighter IMF.

As we demonstrate and discuss below, (semi)-analytical models and
hydrodynamical simulations of galaxy formation do generally not
predict values of $\rj$ that are independent of halo mass, and if they
do, the corresponding values of $\egf$ are inconsistent with the
empirical results inferred here.

\begin{table*}
 \centering
 \caption{Relations between mass and angular momentum of galaxies and dark matter haloes for model A, as presented in Figs.~\ref{fig:jmvir}-\ref{fig:spin}. Uncertainties on halo masses are $1$ and $2\sigma$, and are propagated to other parameters as appropriate. }
  \begin{tabular}{lccccccccc}
\hline
\hline  
$\log_{10}(\Mgal)$ & $\log_{10}(\Mvir)$ & $\egf$ & $\log_{10}(\Jgal/\Mgal)$ &  $\rj$ & $\lampgal$ \\
$[\Msun]$         &  $[\Msun]$         &        & $[\rm kpc \kms]$        &        &            \\
\hline
 9.484 & 11.318$^{+0.125}_{-0.125}$ $^{+0.250}_{-0.250}$ &  0.09 $^{-0.03}_{+ 0.02}$  $^{-0.07}_{+ 0.04}$ & 2.417& 0.57 $^{-0.10}_{+ 0.12}$ $^{-0.18}_{+ 0.27}$ & 0.018$^{-0.003}_{+0.004}$ $^{-0.006}_{+0.008}$\\
 9.631 & 11.414$^{+0.125}_{-0.125}$ $^{+0.250}_{-0.250}$ &  0.10 $^{-0.03}_{+ 0.03}$  $^{-0.08}_{+ 0.04}$ & 2.482& 0.57 $^{-0.10}_{+ 0.12}$ $^{-0.18}_{+ 0.27}$ & 0.018$^{-0.003}_{+0.004}$ $^{-0.006}_{+0.008}$\\
 9.784 & 11.513$^{+0.125}_{-0.125}$ $^{+0.250}_{-0.250}$ &  0.11 $^{-0.04}_{+ 0.03}$  $^{-0.09}_{+ 0.05}$ & 2.550& 0.57 $^{-0.10}_{+ 0.12}$ $^{-0.18}_{+ 0.27}$ & 0.018$^{-0.003}_{+0.004}$ $^{-0.006}_{+0.008}$\\
 9.941 & 11.615$^{+0.125}_{-0.125}$ $^{+0.250}_{-0.250}$ &  0.13 $^{-0.04}_{+ 0.03}$  $^{-0.10}_{+ 0.06}$ & 2.621& 0.58 $^{-0.10}_{+ 0.12}$ $^{-0.18}_{+ 0.27}$ & 0.018$^{-0.003}_{+0.004}$ $^{-0.006}_{+0.008}$\\
10.103 & 11.720$^{+0.125}_{-0.125}$ $^{+0.250}_{-0.250}$ &  0.15 $^{-0.05}_{+ 0.04}$  $^{-0.12}_{+ 0.06}$ & 2.697& 0.59 $^{-0.10}_{+ 0.12}$ $^{-0.19}_{+ 0.27}$ & 0.018$^{-0.003}_{+0.004}$ $^{-0.006}_{+0.008}$\\
10.270 & 11.832$^{+0.125}_{-0.125}$ $^{+0.250}_{-0.250}$ &  0.17 $^{-0.06}_{+ 0.04}$  $^{-0.13}_{+ 0.07}$ & 2.779& 0.60 $^{-0.10}_{+ 0.13}$ $^{-0.19}_{+ 0.28}$ & 0.018$^{-0.003}_{+0.004}$ $^{-0.006}_{+0.009}$\\
10.442 & 11.950$^{+0.125}_{-0.125}$ $^{+0.250}_{-0.250}$ &  0.19 $^{-0.06}_{+ 0.05}$  $^{-0.15}_{+ 0.08}$ & 2.868& 0.61 $^{-0.11}_{+ 0.13}$ $^{-0.19}_{+ 0.29}$ & 0.019$^{-0.003}_{+0.004}$ $^{-0.006}_{+0.009}$\\
10.618 & 12.078$^{+0.125}_{-0.125}$ $^{+0.250}_{-0.250}$ &  0.21 $^{-0.07}_{+ 0.05}$  $^{-0.17}_{+ 0.09}$ & 2.960& 0.62 $^{-0.11}_{+ 0.13}$ $^{-0.20}_{+ 0.29}$ & 0.019$^{-0.003}_{+0.004}$ $^{-0.006}_{+0.009}$\\
10.796 & 12.215$^{+0.125}_{-0.125}$ $^{+0.250}_{-0.250}$ &  0.23 $^{-0.08}_{+ 0.06}$  $^{-0.18}_{+ 0.10}$ & 3.050& 0.62 $^{-0.11}_{+ 0.13}$ $^{-0.20}_{+ 0.29}$ & 0.019$^{-0.003}_{+0.004}$ $^{-0.006}_{+0.009}$\\
10.979 & 12.361$^{+0.125}_{-0.125}$ $^{+0.250}_{-0.250}$ &  0.25 $^{-0.08}_{+ 0.06}$  $^{-0.20}_{+ 0.11}$ & 3.154& 0.63 $^{-0.11}_{+ 0.13}$ $^{-0.20}_{+ 0.29}$ & 0.019$^{-0.003}_{+0.004}$ $^{-0.006}_{+0.009}$\\
11.164 & 12.517$^{+0.125}_{-0.125}$ $^{+0.250}_{-0.250}$ &  0.27 $^{-0.09}_{+ 0.07}$  $^{-0.21}_{+ 0.12}$ & 3.261& 0.63 $^{-0.11}_{+ 0.13}$ $^{-0.20}_{+ 0.30}$ & 0.020$^{-0.003}_{+0.004}$ $^{-0.006}_{+0.009}$\\
11.352 & 12.680$^{+0.125}_{-0.125}$ $^{+0.250}_{-0.250}$ &  0.29 $^{-0.10}_{+ 0.07}$  $^{-0.22}_{+ 0.13}$ & 3.391& 0.66 $^{-0.12}_{+ 0.14}$ $^{-0.21}_{+ 0.31}$ & 0.021$^{-0.004}_{+0.004}$ $^{-0.007}_{+0.010}$\\
\hline
\label{tab:data}
\end{tabular}
\end{table*}

\subsection{Galaxy Spin Parameter}

\begin{figure}
\centerline{
\psfig{figure=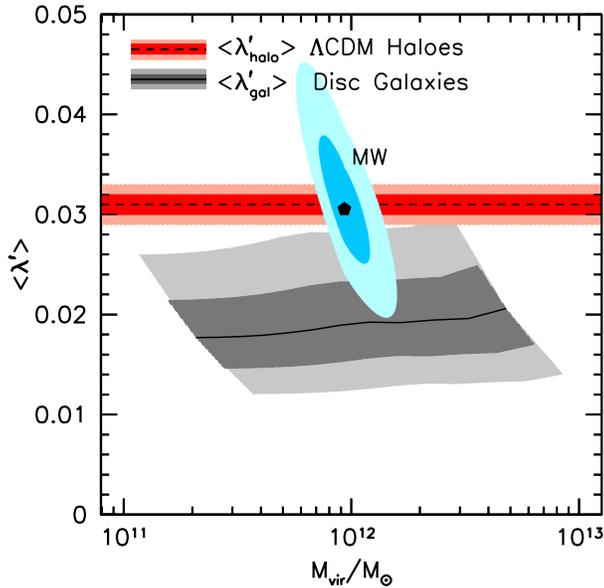,width=0.47\textwidth}
}
\caption{Average spin parameter {\it vs.} halo virial mass, for disc
  galaxies (solid line and grey shaded region), and \LCDM dark matter
  haloes (dashed line, red shaded region). For the galaxy spin
  parameter the shaded regions show the one and two $\sigma$
  uncertainties introduced by the systematic uncertainty in halo
  masses. The pentagon shows an estimate for the Milky Way (MW)
    with the shaded region showing one and two $\sigma$
    uncertainties.}
\label{fig:spin}
\end{figure}

We now discuss the galaxy spin parameter, as this is a quantity that
has been measured and interpreted by previous studies.
Fig.~\ref{fig:spin} shows the average galaxy spin parameter as a
function of halo virial mass. This is the same as the right-hand panel
of Fig.~\ref{fig:jmvir}, but with a zero point shift.  As discussed
above, there is a slight mass dependence to $\lampgal$, but given the
uncertainties in halo masses, $\lampgal$ is consistent with being
independent of halo mass.  The normalization is $\simeq 60\%$ lower
than predicted for \LCDM haloes (dashed line). There are two simple
interpretations of this result. One is that the baryons that form disc
galaxies have lost about 40\% of their angular momentum during galaxy
formation, which could occur via dynamical friction. Another is that
the baryons that form disc galaxies have 40\% lower specific angular
momentum than the cosmically available baryons, which could occur due
to feedback and/or inefficient cooling. We discuss these
interpretations in more detail in \S~\ref{sec:theory}.

An interpretation advocated by Tonini \etal (2006) is that if one
restricts attention to haloes that have not experienced a major merger
since $z \sim 3$, the average spin parameters $\lambda'$ turns out to
be around 0.023 (D'Onghia \& Burkert 2004).  This is motivated by the
idea that bulge-less disc galaxies require a quiescent merger
history. However, this simple expectation has been cast into
  doubt by numerical simulations which have shown that (gas rich)
  discs can survive major mergers (e.g., Springel \& Hernquist 2005;
  Hopkins \etal 2009).  While we cannot rule out a merger history bias
  for massive disc galaxies, we argue that such a bias cannot occur
  for low mass galaxies ($\Mgal \lta 3\times 10^{10}\Msun$, $\Mvir
  \lta 10^{12}\Msun$) because typical low mass galaxies are
  bulge-less, and thus must form in haloes with typical merger
  histories (and hence typical halo spins).

\subsection{Comparison with the Milky Way} 
As a reference, the black pentagons and light blue shaded regions in
Figs.~\ref{fig:jmvir} \& \ref{fig:mmvir} show estimates for the
specific angular momentum and mass of the Milky Way (MW). Here we have
estimated the specific angular momentum of the MW assuming
$\Jgal/\Mgal=2 R_{\rm d} V_{\rm rot}$, which is appropriate for an
exponential disc with a flat rotation curve.  We adopt $R_{\rm d}=2.8
\pm 0.23 \,\kpc (1\sigma)$ based on the dynamical models of Widrow
\etal (2008) and $V_{\rm rot}=219\pm20 \,\kms (1\sigma)$ from the
observations of Reid \etal (1999).  For the baryonic mass of the MW we
use the bulge and disc masses, again from the dynamical models of
Widrow \etal (2008) to obtain $\log_{10}(\Mgal/\Msun)=10.71\pm0.075$
$(1\sigma)$.  For the virial mass of the MW we average the results of
Smith \etal (2007) and Xue \etal (2008) yielding
$\log_{10}(\Mvir/\Msun)=11.97\pm0.22$ $(2\sigma)$.

The resulting galaxy spin parameter (see Eq.2) and galaxy formation
efficiency (see Eq.7) of the MW are $\lampgal=0.031^{+0.016}_{-0.011}$
(2$\sigma$), and $\egf=0.033^{+0.28}_{-0.15}$ $(2\sigma)$,
respectively. Both $\lampgal$ and $\egf$ are higher than for typical
late-type galaxies. This could signify that the MW is atypical, or it
could simply be a reflection of the measurement errors (especially in
halo masses) both for the MW and for samples of late-type galaxies.

\subsection{Comparison with Previous Studies}
\label{sec:previous}

Several studies in the past used similar methods to determine the spin
parameters of disc galaxies. These studies all find higher spin
parameters than we find here, but as we show the differences can be
ascribed to differences in the estimation of halo masses, which is the
dominant source of systematic uncertainty in measuring the spin
parameter.

\begin{figure}
\centerline{
\psfig{figure=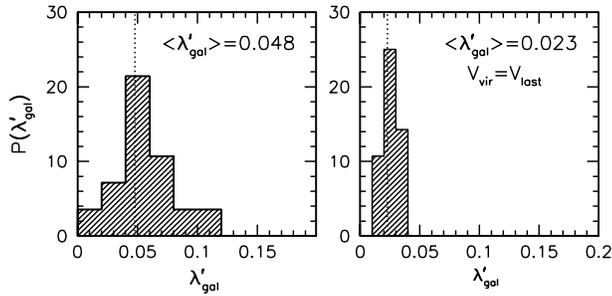,width=0.47\textwidth}
}
\caption{Histograms (hatched) of the distribution of galaxy spin
  parameters ($\lampgal$) for the 14 disc dominated dwarf galaxies
  from van den Bosch, Burkert, \& Swaters (2001; hereafter BBS01).  In
  the left-hand panel $\lampgal$ has been obtained using the halo
  virial velocities from BBS01. In the right-hand panel, the halo
  virial velocities have been calculated assuming $\Vvir=\Vlast$,
  where $\Vlast$ is the rotation velocity at the last measured point
  (i.e., at the largest radius). The latter results in spin parameters
  in good agreement with our results (cf. Fig.~\ref{fig:spin}).}
\label{fig:bbs}
\end{figure}

Van den Bosch, Burkert \& Swaters (2001; hereafter BBS01) used
Eq.~(\ref{eq:jgal}) and halo masses inferred from rotation curve
modelling to determine the galaxy spin parameters for a sample of 14
disc-dominated dwarf galaxies. They found that $\langle \lamp_{\rm
  gal} \rangle = 0.048$ (see Fig.~\ref{fig:bbs})\footnote{Here we have
  converted the results from BBS01 to the spin parameter definition of
  Bullock \etal (2001).}, albeit with a relatively large uncertainty
due to the small sample size and the uncertainties in the stellar
mass-to-light ratios and halo masses.  At first sight, this is much
larger than the average galaxy spin parameters inferred here; $\langle
\lamp_{\rm gal} \rangle = 0.019$. However, there are several factors
that can account for this large difference.  BBS01 used dark halo
masses and sizes derived from fits to rotation curves assuming
adiabatic contraction. Since the rotation curves only extend to $\sim
10-20\%$ of the halo virial radius, this involves a significant
extrapolation, and thus the potential for systematic errors. As
discussed in detail in Dutton \etal (2007), the assumption that disc
formation results in adiabatic contraction of its dark matter halo may
not be realistic.  The galaxies in BBS01 have a median
$\Vlast/\Vvir=1.55$, where $\Vlast$ is the rotation velocity at the
last measured point (i.e., at the largest radius) . Since dark matter
haloes in the \LCDM concordance cosmology have $\Vmax/\Vvir \lta 1.2$,
where $\Vmax$ is the maximum circular velocity, the derived values
from BBS01 are {\it strongly} influenced by the assumption of
adiabatic contraction.  Dutton \etal (2010b) have shown that the halo
virial velocity is, on average, approximately equal to the optical
rotation velocity. Using this assumption to derive halo masses results
in $\log_{10}(\Mvir/\Msun) =11.2\pm0.3$ and a median spin parameter
$\lambda^{\prime}_{\rm gal}=0.023$ (right-hand panel of
Fig.~\ref{fig:bbs}).  If we consider a 10\% uncertainty on
$\Vrot/\Vvir$ (Dutton \etal 2010b) then this results in a range of
$0.019 - 0.028$ for $\lambda^{\prime}_{\rm gal}$, which is consistent
with our measured values (see Fig.~\ref{fig:spin}).

Tonini \etal (2006) used average scaling relations between stellar
mass, gas mass, halo mass and disc size to compute an average galaxy
spin parameter as function of halo mass, under the assumption that
dark matter haloes follow a Burkert (1995) profile. They found that
$\langle \lamp_{\rm gal} \rangle \simeq 0.025-0.030$ with no
significant dependence on halo mass. A comparison between the halo
mass - stellar mass relation used by Tonini \etal (2006) and by
ourselves indicates that their halo masses are a factor of $\sim 2$
lower than ours. This difference in halo masses fully accounts for
their higher spin parameters.

Finally, Hernandez \etal (2007) determined the galaxy spin parameters
for a large sample of galaxies taken from the Sloan Digital Sky Survey
(SDSS). Rather than Tonini \etal (2006), they obtained estimates for
$\lambda_{\rm gal}$ for {\it individual} galaxies using the method
proposed by Hernandez \& Cervantes-Sodi (2006), which is based on the
following assumptions: (i) dark matter haloes are singular isothermal
spheres, (ii) the self-gravity of the disc can be ignored, and (iii)
all galaxies have the same ratio of dark matter mass to disc mass. In
addition, they inferred the rotation velocity of each individual
galaxy in a statistical sense, using an empirical Tully-Fisher
relation. They found that for disc galaxies $\lamp_{\rm gal}$ follows
a log-normal distribution with $\langle \lamp_{\rm gal} \rangle \sim
0.059$, which is a factor of $\sim 3$ higher than our result.  The
method used by Hernandez \etal (2007) makes several highly
questionable assumptions, and thus we do not consider their result
reliable.  For example, Hernandez \etal (2007) assume a constant
galaxy mass to dark matter mass ratio of $F=0.04$, which corresponds to a
galaxy formation efficiency of $\egf \simeq 0.25$. Comparison with
Fig.~\ref{fig:mmvir} shows that while this is a good assumption for a
$2\times 10^{12}\Msun$ halo, it significantly over-estimates the
galaxy formation efficiency, and hence spin parameters, of lower mass
galaxies. Since Hernandez \etal (2007) use a volume limited sample,
their measurement of the spin parameter is biased towards lower mass
galaxies, where their assumption of $F=0.04$ is high by a factor of
$\simeq 2$.  In the method used by Hernandez \etal (2007) $\lamp_{\rm
  gal}\propto F$, and thus their assumption of $F=0.04$ accounts for
at least half the discrepancy between our result and theirs.

\begin{figure}
\centerline{
\psfig{figure=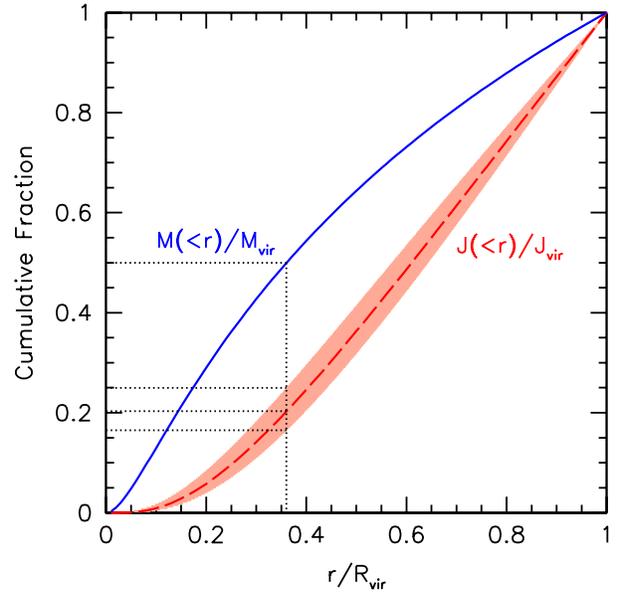,width=0.47\textwidth}
}
\caption{The cumulative, normalized fractions of mass (blue line) and
  angular momentum (dashed red line) for \LCDM haloes (Navarro \etal
  1997; Bullock \etal 2001). The shaded region shows the uncertainty
  on $J(r)/\Jvir$ assuming $s=1.3\pm0.3$ in Eq.~\ref{eq:bul}. This
  clearly demonstrates that mass is more centrally concentrated than
  angular momentum. For example, the inner 50\% of the mass contains
  just $20^{+5}_{-4}\%$ of the angular momentum (black dotted
  lines). Thus if the efficiency of galaxy formation is regulated by
  the efficiency of cooling, we would expect a strong correlation
  between galaxy spin parameter and galaxy formation efficiency, which
  is not observed.}
\label{fig:jdmd}
\end{figure}

\section{Comparison with Galaxy Formation Models}
\label{sec:theory}

\subsection{Inside-Out Cooling Models}
\label{sec:cooling}

If the galaxy formation efficiency is determined by the efficiency of
cooling, then we expect that $\rj$ will be higher for higher galaxy
formation efficiencies. This is because in \LCDM haloes mass is more
centrally concentrated than angular momentum.  This is illustrated in
Fig.~\ref{fig:jdmd}. The blue solid line shows the cumulative mass
fraction, $M(<r)/M_{\rm vir}$ as a function of radius for a NFW halo
with concentration $c=10$.  The red dashed line shows the cumulative
angular momentum fraction, $J(<r)/J_{\rm vir}$, as a function of
radius assuming
\begin{equation}
\label{eq:bul}
 \frac{J(<r)/\Jvir}{M(<r)/\Mvir} = [M(<r)/\Mvir]^{s},
\end{equation}
where $s=1.3\pm0.3$ (Bullock \etal 2001).
Under the assumptions that (i) the baryons initially acquire the same
specific angular momentum as the dark matter, and (ii) the baryons
accrete onto the central galaxy inside-out, then Eq.~\ref{eq:bul}
implies $\rj = \egf^{s}$.

\begin{figure}
\centerline{
\psfig{figure=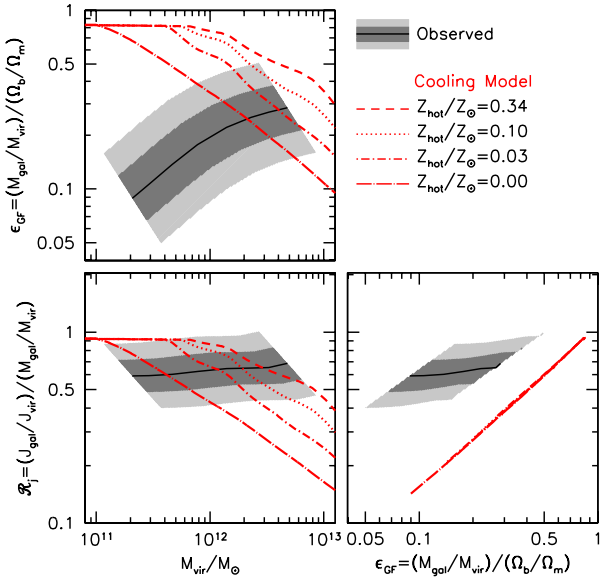,width=0.47\textwidth}
}
\caption{Correlations between halo mass, $\Mvir$, galaxy formation
  efficiency, $\egf$, and specific angular momentum ratio, $\rj$, for
  models and observations at redshift $z=0$. The empirical relations
  for disc galaxies are shown as solid black lines, with the shading
  reflecting the one and two $\sigma$ uncertainties on halo
  masses. Note that the galaxy formation efficiency increases with
  increasing halo mass, while the specific angular momentum ratio is
  independent of halo mass.  The corresponding relations for a simple
  cooling model are shown with red lines, where the different line
  types correspond to different metallicities, as indicated. In this
  cooling model $\rj \sim \egf$ (independent of the metallicity of the
  hot gas), inconsistent with the empirical findings.}
\label{fig:erm}
\end{figure}

We now extend this simple inside-out cooling model to cosmologically
evolving dark matter haloes.  The red lines in Fig.~\ref{fig:erm} show
the relations between $\rj$, $\egf$ and $\Mvir$ for a scenario in
which the galaxy formation efficiency is determined by the efficiency
of cooling. This model is a simplified version of the galaxy formation
model from Dutton \& van den Bosch (2009). Briefly this model consists
of NFW haloes that grow by smooth accretion of baryons and dark
matter. The angular momentum of each shell of accreted gas is
determined by assuming that the angular momentum distribution (AMD) of
a given halo is independent of time. The AMD is specified by two
parameters: the normalization ($\lambda$) and shape ($\alpha$).  When
the baryons enter the halo they are shock heated to the virial
temperature. The cooling time depends on the temperature, density and
metallicity of the hot gas. We use the metallicity dependent
collisional ionization equilibrium cooling functions of Sutherland \&
Dopita (1993).  Results for four different metallicities from
primordial to one third solar are shown.  For low mass haloes
($\Mvir\sim 10^{11}\Msun$) cooling is very efficient, so that $\simeq
90\%$ of the baryonic mass is accreted (left-hand panels)\footnote{It
  is not 100\% because it takes a free fall time for the most recently
  accreted baryons to reach the galaxy.}. For more massive haloes, the
galaxy formation efficiency, $\egf$, depends strongly on the
metallicity of the gas, with higher metallicities yielding larger
values of $\egf$. Clearly, the trend that $\egf$ decreases with
increasing halo mass is completely opposite to the empirical
result. As for the parameter $\rj$, the simple cooling models predict
that $\rj$ should decrease with increasing halo mass, which is a
consequence of cooling being an inside-out process combined with the
fact that mass is more centrally concentrated than angular momentum
(cf.  Fig.~\ref{fig:jdmd}). Hence, in massive haloes where $\egf \ll
1$ due to inefficient cooling, the galaxy ends up having less specific
angular momentum than its dark matter halo. This strong mass
dependence of $\rj$ is again inconsistent with our empirical
results. Finally, the cooling models predicts an almost linear
relation between $\rj$ and $\egf$, which is independent of the gas
metallicity (lower right-hand panel of Fig.~\ref{fig:erm}).  Clearly,
this linear relation is very different from its empirical equivalent,
indicating that additional mechanisms for accreting and/or expelling
gas are required.

\begin{figure}
\centerline{
\psfig{figure=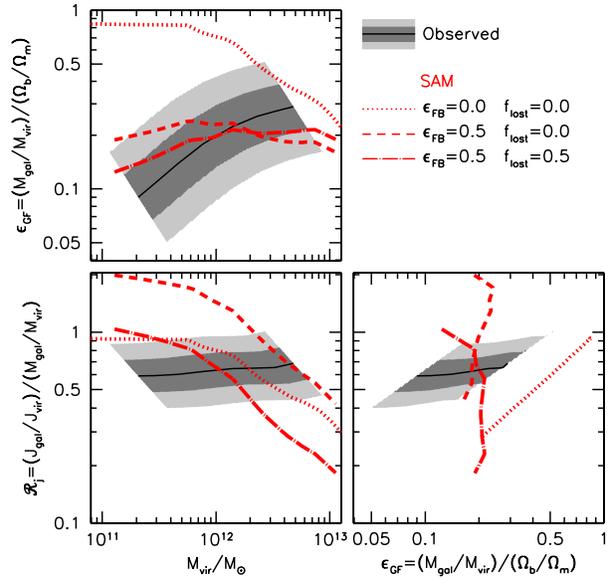,width=0.47\textwidth}
}
\caption{Same as Fig.~\ref{fig:erm}, but this time we compare the
  empirical relations (solid black lines with shaded regions) with
  models that include feedback and/or angular momentum loss (red
  lines). All models shown adopt a hot gas metallicity of $Z_{\rm hot}
  = 0.1 Z_{\odot}$, but they differ in feedback efficiency, $\efb$,
  and the fraction of angular momentum lost by the baryons during
  galaxy formation, $\flost$. The dotted line ($\efb = \flost = 0$) is
  identical to the simple cooling model from Fig.~\ref{fig:erm}, and
  is shown for comparison. Note that none of the models is able to
  simultaneously match the low galaxy formation efficiencies and the
  constant angular momentum ratio. See text for a detailed discussion.}
\label{fig:erm2}
\end{figure}

\subsection{Outflows and Angular Momentum Loss}
\label{sec:outflows}

We now consider two modifications to the simple cooling model:
outflows and angular momentum loss. We assume that outflows, which
move at the local escape velocity, are driven by energy from
supernovae (SNe). The fraction of the SNe energy that goes into an
outflow is specified by the free parameter $\efb$ (see van den Bosch
2002; Dutton \& van den Bosch 2009). Other than inefficient cooling in
massive haloes, and UV heating in low mass haloes, galactic outflows
are the primary mechanism for explaining the observed low galaxy
formation efficiencies (e.g., Dekel \& Silk 1986; Cole \etal 1994).
For simplicity, we assume that outflow gas leaves the disc and halo,
and does not return. However, recent cosmological hydrodynamical
simulations have shown that ejected gas may return to the galaxy
(Oppenheimer \etal 2010).  Some of the ejected gas may mix with the
halo gas and thus re-accrete with higher specific angular momentum
(Brook \etal 2012). Since this process is a re-distribution (i.e.,
ejected gas gains angular momentum, halo gas loses angular momentum),
it does not change the specific angular momentum of the baryons.  If
the halo gas accretes onto the central galaxy (as is expected for low
mass, $\Mvir\lta 10^{12}\Msun$, haloes) then there is no net change in
the angular momentum of the galaxy. If the halo gas does not accrete
(which is expected for high mass haloes) then the galaxy can gain
angular momentum via this galactic fountain effect. Thus if we were to
implement re-accretion into our model we would need higher feedback
efficiencies to achieve the same galaxy formation efficiencies, but to
first order there would be no change to the specific angular momentum.

We assume that during the accretion process, a fraction $\flost$ of
the angular momentum of the gas is lost to the dark matter halo.  The
mechanism responsible for this transfer is dynamical friction. In our
models we do not take into account the effects of angular momentum
transfer and outflows on the structure of the dark matter halo. But we
note that both of these processes will result in halo expansion, or at
least a reduction of the amount of adiabatic contraction, i.e., a
lower effective value of $\nu$ (e.g., Navarro \etal 1996; El-Zant
\etal 2001; Mo \& Mao 2004; Cole, Dehnen \& Wilkinson 2011)

\begin{figure}
\centerline{
\psfig{figure=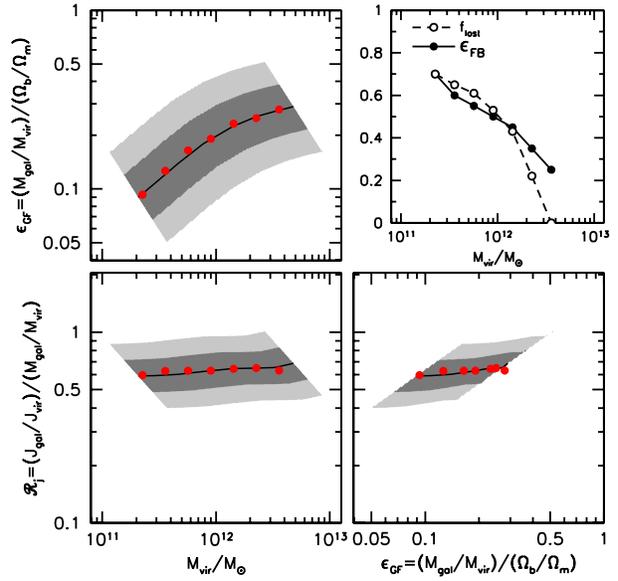,width=0.47\textwidth}
}
\caption{Same as Fig.~\ref{fig:erm2}, but this time we have tuned the
  feedback efficiency, $\efb$, and angular momentum loss, $\flost$, of
  the model (red, solid dots) in order to accurately reproduce the
  empirical relations (solid black lines with shaded regions). The
  small, upper right-hand panel shows the corresponding halo mass
  dependence of $\efb$ (solid circles) and $\flost$ (open circles):
  reproducing the empirical relations requires both of these
  parameters to decrease strongly with increasing halo mass. See text
  for a detailed discussion.}
\label{fig:erm3}
\end{figure}

\begin{figure}
  \centerline{
\psfig{figure=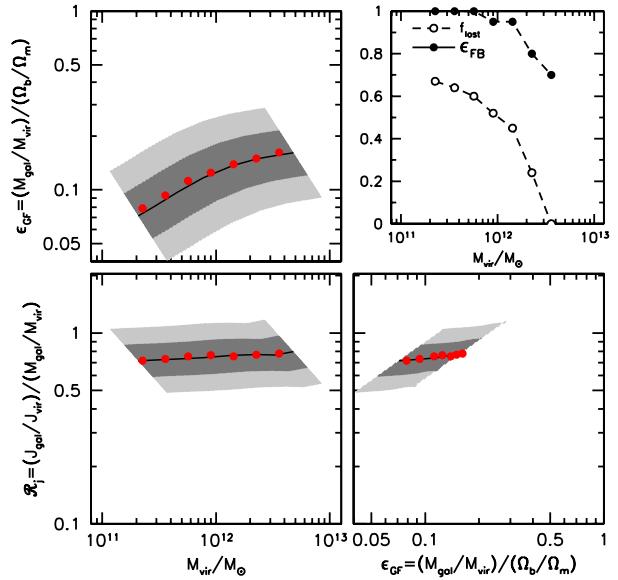,width=0.47\textwidth}
  }
  \caption{Same as Fig.~\ref{fig:erm3}, but here we have used model~B
    (stellar mass-to-light ratios that are 0.3 dex lower and halo
    contraction characterized by $\nu = 0.8$) for the empirical
    relations.  Reproducing the inferred galaxy formation efficiencies
    now requires even stronger feedback (i.e., larger $\efb$), while
    the fraction of angular momentum that is lost is similar to that
    for our fiducial model shown in Fig.~\ref{fig:erm3}.}
\label{fig:erm4}
\end{figure}

In Fig.~\ref{fig:erm2} we show several models with different values of
$\efb$ and $\flost$. In all models we assume that halo gas is enriched
to one tenth solar. The dotted line shows a model with $\efb=0.0$ and
$\flost=0.0$: this corresponds to the simple cooling model from
Fig.~\ref{fig:erm} and is shown for comparison. The galaxy formation
efficiencies are too high for all halo masses.  In order to produce
realistic galaxy formation efficiencies outflows are needed. A model
with $\efb=0.5$ and $\flost=0.0$ (short dashed line) has
$\egf\sim20\%$.  In our model outflows increase $\rj$ by
preferentially removing gas with low specific angular momentum. Since
outflows are driven by star formation, regions that have higher star
formation rates have higher mass outflow rates.  Star formation is
empirically more efficient at smaller galactic radii (Kennicutt 1998)
as well as at higher redshifts (e.g., Noeske \etal 2007; Daddi \etal
2007; Elbaz \etal 2007). Smaller galactic radii correspond to lower
specific angular momentum and discs are smaller at higher redshifts
(e.g., Dutton \etal 2011a); both of these effects contribute to
outflows increasing the specific angular momentum of the material that
remains in the galaxy.  In our energy driven outflow model the mass
loading factor ($\equiv$ outflow rate / star formation rate) is higher
in lower mass galaxies, which results in $\rj$ increasing more in
lower mass haloes (by up to a factor two). This increases the
disagreement with the empirically inferred relation between $\rj$ and
halo mass. We can lower $\rj$ by introducing angular momentum loss
(i.e., setting $\flost > 0$). The long dashed-dotted lines in
Fig.~\ref{fig:erm2} correspond to a model with $\efb=0.5$ and
$\flost=0.5$. This angular momentum loss reduces $\rj$ by a factor
$\sim 2$ at all mass scales, bringing it in reasonable agreement with
the observed value for haloes with $\Mvir\simeq
10^{12}\Msun$. However, for lower (higher) mass haloes the model
overpredicts (underpredicts) the value of $\rj$ compared to the
empirically inferred values.  Increasing $\flost$ also modified the
galaxy formation efficiency because it results in denser (smaller)
discs.  In low mass haloes, the galaxies are gas rich. Hence, an
increase in disc density results in higher star formation rates and
consequently more energy to drive outflows. This more than compensates
for the higher escape velocities, resulting in lower values of
$\egf$. In massive haloes, on the other hand, the galaxies are gas
poor, so that an increase of the density of the disc only results in
an increase of the escape velocity. This reduces the amount of ejected
gas, and therefore increases $\egf$.

It is clear that none of the models discussed thus far can
simultaneously reproduce the empirically inferred trends of $\egf$ and
$\rj$ with halo mass. As an illustration of what seems to be required,
we now consider a model in which both $\efb$ and $\flost$ depend on
halo mass. The red filled symbols in Fig.~\ref{fig:erm3} show a model
in which \efb and \flost have been tuned so that the model matches the
empirical $\egf$ and $\rj$. The corresponding values of $\efb$ and
$\flost$, as a function of virial mass, are shown in the upper
right-hand panel. Lower mass haloes {\it require} higher $\efb$ and
$\flost$. For example, this model has $(\efb,\flost)=(0.7,0.7)$ for
$\Mvir\simeq 10^{11.3}\Msun$, and $(\efb,\flost)=(0.25,0.0)$ for
$\Mvir\simeq 10^{12.4}\Msun$.

The results shown in Fig.~\ref{fig:erm3} correspond to the empirical
$\egf$ and $\rj$ inferred under the assumptions of model~A (i.e.,
Chabrier IMF and halo contraction characterized by $\nu = -0.5$).
Fig.~\ref{fig:erm4} shows the analogous results but for model~B (an
IMF that results in stellar mass-to-light ratios that are 0.3 dex
lower than for model A and with $\nu = 0.8$).  Reproducing the galaxy
formation efficiencies now requires even stronger feedback, but with a
similar mass dependence, whereas the fraction of angular momentum that
is lost is similar to that of our fiducial model.

\begin{figure}
\centerline{
\psfig{figure=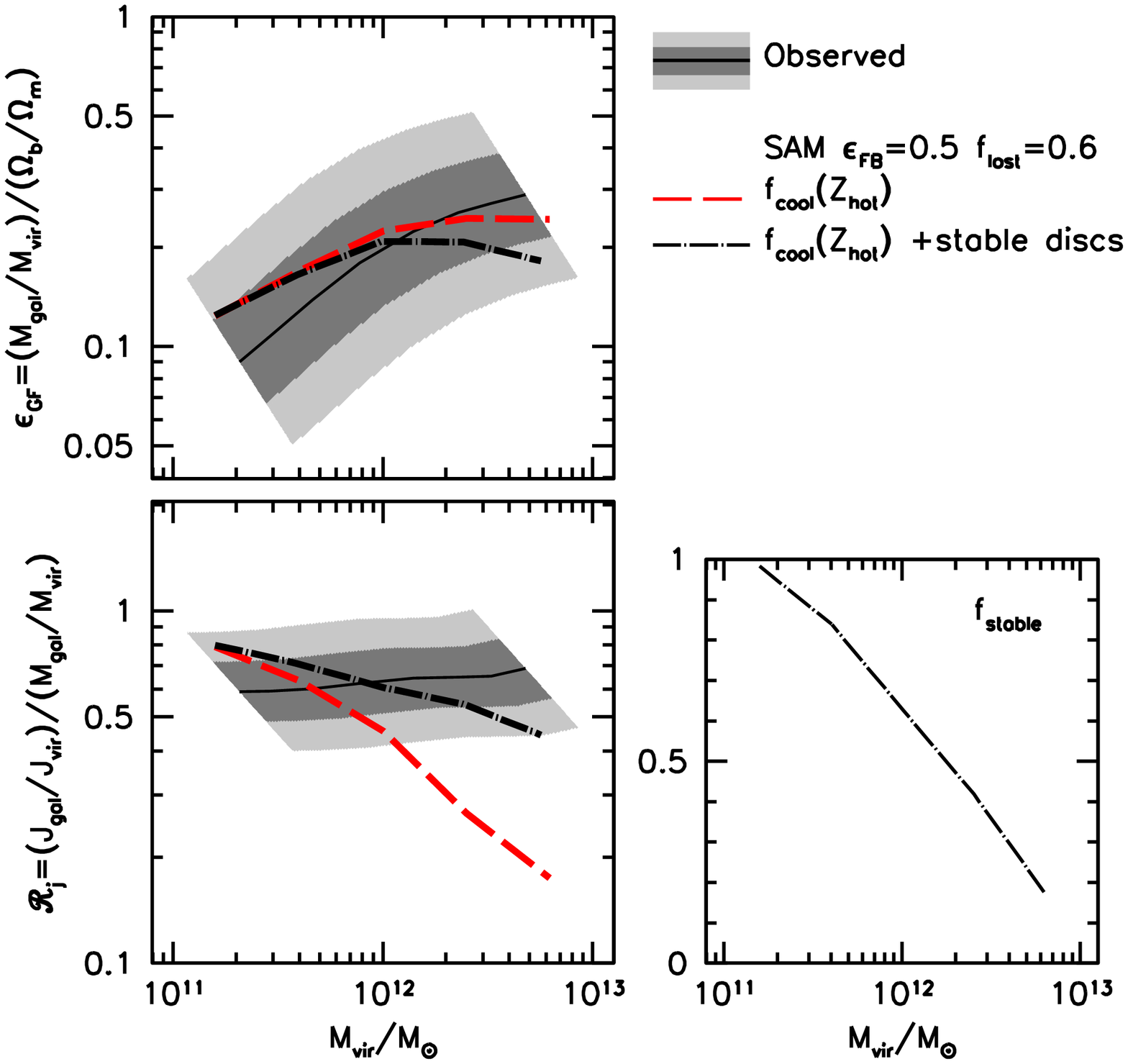,width=0.47\textwidth}
}
\caption{The impact of disc stability on the galaxy formation
  efficiency ($\egf$; upper left-hand panel) and specific angular
  momentum ratio ($\rj$; lower let-hand panel). The solid line with
  the shaded regions indicate the empirical relations (for our
  fiducial model~A). The dashed, red line corresponds to a simple model
  with $Z_{\rm hot} = 0.1 Z_{\odot}$, $\efb = 0.5$ and $\flost = 0.6$,
  which yields galaxy formation efficiencies in reasonable agreement
  with the data. However, it predicts that $\rj$ decreases strongly
  with increasing halo mass, in clear conflict with the data. The
  dot-dashed, black line corresponds to the same model, but now we
  have removed those discs that do not satisfy our disc stability
  criterion. This removes massive galaxies in low-spin haloes,
  resulting a higher $\rj$ for mass haloes, in better agreement with
  the data.}
\label{fig:erm5}
\end{figure}

\subsection{Disc Stability}

Thus far we have assumed that disc galaxies form in typical dark
matter haloes. This must be true at the low mass end, simply because
disc galaxies dominate the galaxy population at low masses. However,
at high masses disc galaxies are a minority, so that it is possible
that they form in a biased subset of dark matter haloes. To reconcile
our models with the empirical fact that $\rj$ seems to be independent
of halo mass requires a bias such that massive disc galaxies form
preferentially in haloes with larger spin parameters.

Such a bias is naturally achieved by invoking disc stability.  Disc
galaxies can only survive as such to the present day if their discs
are sufficiently stable. Disc galaxies that are unstable redistribute
their angular momentum via secular evolution, typically resulting in
spheroid dominated galaxies (e.g., Combes \etal 1990; Norman, Sellwood
\& Hasan 1996; van den Bosch 1998; Mao \& Mo 1998).  Since more
self-gravitating (denser) discs are more unstable (e.g., Efstathiou,
Lake \& Negroponte 1982), and since the surface density of a disc is
proportional to $\lambda_{\rm halo}^{-2}$ (e.g., Mo, Mao \& White
1998), one expects a bias against disc galaxies in haloes with the
lowest spin parameter. Since the disc's surface density is also
proportional to $\egf$, and empirically $\egf$ increases with
increasing halo mass, one expects discs in less massive haloes to be
more stable than those in more massive haloes. Hence, disc stability
may introduce a bias against forming {\it massive} disc galaxies in
{\it low} spin parameter haloes, exactly what seems to be required.

To test the potential impact of disc stability on the relations
between $\egf$, $\rj$, and $\Mvir$, we add a simple stability
criterion to our models: we assume that if the disc contributes more
than 84\% of the mass within 2.2 disc scale lengths it will be highly
unstable, and we remove these galaxies from our sample of model
galaxies.  We note that discs are already expected to become unstable
--- and hence to form bulges via secular evolution --- at disc mass
fractions lower than 84\%. However, we choose a high stability
threshold as we do not wish to remove the galaxies that remain disc
dominated after secular evolution.

Fig.~\ref{fig:erm5} shows the effect of disc stability on $\egf$ and
$\rj$. In our fiducial model (red long-dashed lines) the fraction of
gas that cools at each time step ($f_{\rm cool}$) depends on the
metallicity of the hot halo gas (here $Z_{\rm hot} = 0.1 Z_{\odot}$),
the feedback efficiency (here $\efb=0.5$) and the fraction of angular
momentum that is lost (here $\flost=0.6$). This model approximately
reproduces the observed relation between galaxy formation efficiency
and halo mass, but it predicts that $\rj$ decreases strongly with
increasing halo mass, which is inconsistent with the empirical
results.  Invoking  the disc stability criteria discussed above
(black dot-long-dashed lines) results in a lower fraction of stable
discs in higher mass haloes (lower right panel): when computing the
average $\rj$ at a given halo mass (for both models and observations),
we assume that the haloes have a median spin parameter as expected
from cosmological simulations (i.e., $\lamp=0.031$). However, since
stable discs have higher median spin parameters than unstable discs,
this results in stable discs having higher $\rj$.  The stable discs in
our model still have an anti-correlation between $\rj$ and halo mass,
but the model is now consistent with the observations at the $2\sigma$
level.

The disc stability scenario also provides a qualitative explanation
for two additional observational facts: (i) the scatter in disc sizes
decreases with increasing stellar mass (Shen \etal 2003), such that
for the most massive discs the scatter is half that expected from the
scatter in halo spin parameters (Dutton \etal 2011a); and (ii) the
fraction of galaxies that are star forming disc galaxies declines with
increasing galaxy and halo mass (e.g., Yang \etal 2008). Hence, we
conclude that disc stability may be an important ingredient for
understanding the empirical scaling relations between $\egf$, $\rj$
and halo mass.

\subsection{Comparison with Hydrodynamical Simulations}
\label{sec:hydro}

The simplistic analytical models used above indicate that it is
not easy to understand why the specific angular momentum ratio
$\rj$ seems to be independent of halo mass, whereas the galaxy
formation efficiency increases strongly with halo mass. In fact, 
reproducing the empirical scaling relations between $\egf$, $\rj$
and halo mass seems to require a feedback efficiency that declines
with increasing halo mass. 

\begin{figure}
\centerline{
\psfig{figure=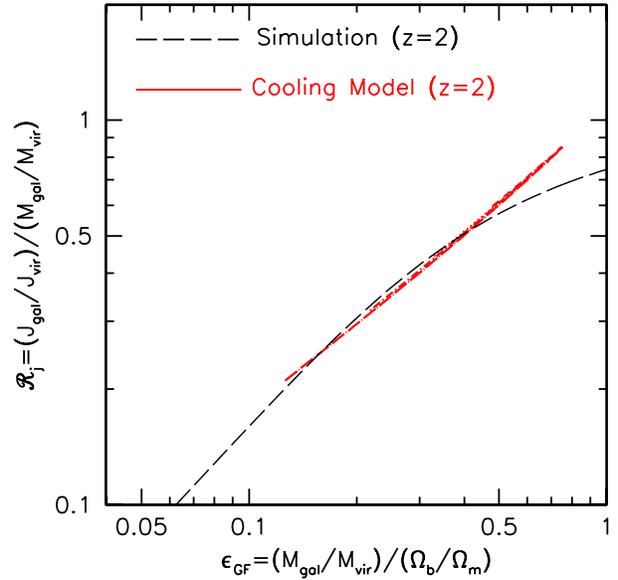,width=0.47\textwidth}
}
\caption{Correlation between specific angular momentum ratio, $\rj$,
  and galaxy formation efficiency, $\egf$. The red lines show the
  relations from our simple inside-out cooling model with different
  gas metallicities (see Fig.~\ref{fig:erm}).  The black, dashed line
  shows the relation from Sales \etal (2009), obtained for simulated
  galaxies at $z\sim 2$ in a cosmological hydrodynamical
  simulation. Note that these simulated galaxies follow exactly the
  relation between $\rj$ and $\egf$ expected for a simple cooling
  model, despite the fact that the simulations include supernova
  feedback. See text for a detailed discussion.}
\label{fig:er}
\end{figure}

An important downside of our simplistic models is that they do not
properly account for the hydrodynamics of outflows from galaxies
embedded in a large scale environment from which the galaxy is also
accreting matter. This requires cosmological, hydrodynamical
simulations, which arguably are the best tool available to model the
complex, hierarchical nature of galaxy formation.  Recently, Sales
\etal (2009) have presented a suite of high resolution, hydrodynamical
simulations of the formation of 
galaxies in a $\Lambda$CDM concordance cosmology, with a wide range of
feedback recipes. Their resulting 
galaxies (at $z=2$) have properties that are well fit by
\begin{equation}\label{eq:sales}
  \rj  = 9.71 \mgal [1 - \exp(-1/9.71\mgal)].
\end{equation}
independent of the feedback recipe!  Here $\mgal=\Mgal/\Mvir=\egf \,
(\Omegab/\Omegam)$ is the galaxy mass fraction.  As a specific
example, for a galaxy mass fraction of $\mgal = 0.033$ (i.e. 20\% of
the cosmic baryon fraction) the specific angular momentum ratio is
predicted to be $\rj = 0.31$.  This scaling is given by the black
dashed line in Fig.~\ref{fig:er}. Interestingly, the relation from
Sales \etal (2009) is very similar to the results from our simple
cooling model (red lines). This suggests that feedback in their
simulations largely preserves the rank order in binding energy of the
baryons, such that the relation between $\egf$ and $\rj$ of the
resulting galaxies is predominantly set by the relative radial
profiles of mass and angular momentum shown in Fig.~\ref{fig:jdmd}.
It is unclear at present why the simulations seem to preserve the rank
order of binding energy, but it may have to do with the fact that
outflows are inhibited from travelling far from their source due to
pressure confinement of the surrounding gas, an effect that is not
accounted for in our simplistic, analytical models. Irrespective of
what causes the relation between $\rj$ and $\egf$ in the simulations,
it is clear that Eq.~\ref{eq:sales} is inconsistent with the empirical
relation.  We caution, though, that the simulations were only run to
redshift $z = 2$, whereas our empirical results correspond to $z\simeq
0$.  In addition, our empirical results are for star-forming
disc-dominated galaxies, whereas the results from the simulations are
for all types of galaxies. It remains to be seen whether simulations
of disc formation that are run to $z=0$ yield a relation between $\rj$
and $\egf$ in better agreement with the data.

\section{Summary}
\label{sec:sum}

We have combined measurements of halo virial masses from weak lensing
and satellite kinematics with the Tully-Fisher and size-mass relations
of disc galaxies (i.e., star-forming and disc-dominated) to infer the
average spin parameter of disc galaxies as a function of their halo
mass.  Using toy models for disc galaxy formation we have discussed
implications of these results for cooling, outflows and angular
momentum loss. We summarize our results as follows:

\begin{itemize}

\item The average galaxy spin parameters of disc galaxies are
  consistent with being independent of halo mass for the range of halo
  masses probed here: $11.3 \leq \log_{10}(\Mvir/\Msun) \leq
  12.7$.

\item The primary uncertainty in measuring the spin parameters of
  galaxies is the determination of halo masses. Realistic
  uncertainties of 0.25 dex ($2\sigma$), result in uncertainties in
  galaxy spin parameters of a factor of 1.5. A secondary uncertainty
  is the stellar IMF. Since gas discs have higher specific angular
  momentum than stellar discs, lower stellar mass normalizations will
  result in higher spin parameters. For example, stellar masses lower
  by 0.3 dex result in spin parameters higher by 0.12 dex.

\item The average spin parameter of galaxies, $\langle \lambda_{\rm
  gal} \rangle = 0.019^{+0.004}_{-0.003}$ ($1\sigma$), is smaller than
  that of their host dark matter haloes, $\langle \lambda_{\rm halo}
  \rangle = 0.031\pm0.001$ (Macci\`o \etal 2007, 2008; Bett \etal
  2007).  The inferred spin parameters of disc galaxies reveal no
    dependence on halo mass, so that the specific angular momentum
    ratio $\rj\simeq 0.6$, independent of halo mass. The galaxy
  formation efficiency parameter, $\egf$, however, reveals a strong
  mass dependence, increasing from $\sim 10\%$ for $\Mvir =
  10^{11.5} \Msun$ to $\sim 30 \%$ for $\Mvir = 10^{12.5} \Msun$
  for our fiducial model.

\item Since mass is more centrally concentrated than angular momentum
  in \LCDM haloes a simple inside-out accretion model results in a
  strong correlation between galaxy spin parameter and galaxy
  formation efficiency, contrary to observations. This provides
  further support for the already popular notion than the low galaxy
  formation efficiencies in haloes of mass $\Mvir \lta 10^{11.7}
  \Msun$ are determined primarily by feedback processes, and not by an
  inefficiency of cooling.

\item The empirically inferred relations between galaxy formation
  efficiency, $\egf$, angular momentum ratio, $\rj$, and halo mass
  seem to indicate that galaxy formation involves galactic outflows,
  or other mechanisms that can regulate the galaxy formation
  efficiency (e.g., supernova feedback, AGN feedback, reionization,
  pre-heating), as well as mechanisms that cause a substantial
  transfer of angular momentum from the baryons to the dark matter
  (e.g., dynamical friction). Although the need for feedback
  mechanisms to regulate the efficiency of galaxy formation has long
  been recognized, the observed scaling relations for disc galaxies
  seem to require that the often adopted supernovae feedback efficiency
  parameter, $\efb$, decreases with halo mass. In addition, the
  angular momentum content of observed disc galaxies also require that
  angular momentum loss is more important for disc formation in less
  massive haloes. It remains to be seen whether realistic models for
  galaxy formation can achieve effective values for $\efb$ and
  $\flost$ that reveal such a dependence on halo mass.

\end{itemize}

What could be the cause for the {\it effective} values for $\efb$ and
$\flost$ to decrease with increasing halo mass? One possibility is
that this mass dependence is a manifestation of the transition from
cold-mode accretion at $\Mvir \lta 10^{11.7}\Msun$ to hot-mode
accretion for haloes with $\Mvir \gta 10^{11.7}\Msun$ (e.g., Birnboim
\& Dekel 2003; Keres \etal 2005, 2009; Brooks \etal 2009). In the cold
mode regime, outflows are unhindered by hot-gas atmospheres, which
may result in higher effective feedback efficiencies. If, as
envisioned in our models, the outflows preferentially remove low
angular momentum material, the same outflows which reduce the galaxy
formation efficiency are expected to increase the specific angular
momentum ratio, $\rj$. Hence, in order to reproduce the inferred $\rj
\simeq 0.6$, the baryons that end up in the disc need to lose a
significant fraction ($\sim 60\%$) of their angular momentum during
the galaxy formation process. If the cold accretion is sufficiently
clumpy, this may come about because of dynamical friction.

In the hot mode accretion regime, outflows are expected to be less
efficient, since they have to do work against their hot gaseous
atmospheres. In this regime, one expects that the baryon fraction and
angular momentum of the resulting disc galaxies are largely determined
by the inside-out cooling of the hot gas, resulting in $\rj$
decreasing strongly with increasing halo mass. This, however, is
inconsistent with the empirical result that $\rj$ appears to be
independent of halo mass.  We have argued that disc stability may play
an important role here, effectively removing systems with low halo
spin parameter from the sample of disc galaxies (because they produce
unstable discs), thereby increasing the average $\rj$ of massive disc
galaxies.  Another important process may be that at high redshifts ($z
\gta 2$), massive haloes in the hot-mode accretion regime can still
accrete a significant amount of material via cold streams that
penetrate the hot halo (e.g., Dekel \etal 2009). Depending on the
`impact parameters' of such streams, the cold material thus deposited
to the central galaxy may have specific angular momentum that is
relatively high, boosting the value of $\rj$ with respect to that
expected in a simple cooling model.

Although plausible, we caution that the above discussion regarding the
potential impact of the cold-mode to hot-mode accretion regime is
highly speculative. In fact, the high resolution, hydrodynamical
simulations of Sales \etal (2009), which in principle should naturally
account for the cold mode {\it vs.} hot mode effects mentioned above,
did not yield (disc) galaxies with the correct relation between $\egf$
and $\rj$. Amazingly enough, their simulated galaxies revealed a
tight, linear correlation between $\egf$ and $\rj$ in perfect
agreement with our predictions for a simple cooling model, {\it
  despite the fact that their simulations included star formation and
  feedback (using several different implementations).}  If taken at
face value, these simulations seem to suggest that feedback has little
to no effect on the $\egf$ {\it vs.} $\rj$ relation; although it
reduces $\egf$, it causes a similar suppression of $\rj$. In addition,
the Sales \etal simulations (which were only run to $z=2$) seem to
indicate that cold mode accretion, which should have been the
dominant mode of accretion for the galaxies in their simulations, does
not have the effect envisioned above. More detailed hydrodynamical
simulations of galaxy formation (run all the way to $z=0$) are needed
to investigate these issues in more detail. In particular, we
currently lack a proper understanding of how cold mode accretion and
cold streams contribute to the build-up of mass and, in particular,
angular momentum of disc galaxies.

\section*{Acknowledgments} 

We thank Julio Navarro for useful discussions.  AAD acknowledges
financial support from the Canadian Institute for Theoretical
Astrophysics (CITA) National Fellows Program.


\label{lastpage}
\end{document}